\documentclass[authoryear,preprint]{elsarticle}
\usepackage[framed,autolinebreaks,useliterate]{mcode}
\usepackage{natbib}
\usepackage{amsmath}
\usepackage{lineno}
\usepackage{rotating}
\usepackage{hyperref}
\usepackage{amssymb}
\usepackage{amsfonts}
\usepackage{graphicx}
\usepackage{todonotes}
\usepackage{pdflscape}
\journal{Neuroimage}
\usepackage{color,transparent}
\begin{document}

\begin{frontmatter}
\title{Multiscale statistical testing for connectome-wide association studies in fMRI}
\author[a,b]{Pierre~Bellec\corref{cor1}}
\ead{pierre.bellec@criugm.qc.ca}
\cortext[cor1]{Corresponding author}
\author[a,c]{Yassine~Benhajali}
\author[d]{Felix~Carbonell}
\author[a,b]{Christian~Dansereau}
\author[a,e]{Genevi\`eve~Albouy}
\author[e]{Maxime~Pelland}
\author[f,g]{Cameron~Craddock}
\author[e]{Olivier~Collignon}
\author[a,e]{Julien~Doyon}
\author[h,i]{Emmanuel~Stip}
\author[a,h]{Pierre~Orban}

\address[a]{Functional Neuroimaging Unit, Centre de Recherche de l'Institut Universitaire de G\'eriatrie de Montr\'eal}
\address[b]{Department of Computer Science and Operations Research, University of Montreal, Montreal, Quebec, Canada}
\address[c]{Department of Anthropology, University of Montreal, Montreal, Quebec, Canada}
\address[d]{Biospective Incorporated, Montreal, Quebec, Canada}
\address[e]{Department of Psychology, University of Montreal, Montreal, Quebec, Canada} 
\address[f]{Nathan Kline Institute for Psychiatric Research, Orangeburg, NY, United States of America}
\address[g]{Center for the Developing Brain, Child Mind Institute, New York, NY, United States of America}
\address[h]{Department of Psychiatry, University of Montreal, Montreal, Quebec, Canada}
\address[i]{Centre Hospitalier de l'Universit\'e de Montr\'eal, Montreal, Quebec, Canada}

\begin{abstract}
Alterations in brain connectivity have been associated with a variety of clinical disorders using functional magnetic resonance imaging (fMRI). We investigated empirically how the number of brain parcels (or scale) impacted the results of a mass univariate general linear model (GLM) on connectomes. The brain parcels used as nodes in the connectome analysis were functionnally defined by a group cluster analysis. We first validated that a classic Benjamini-Hochberg procedure with parametric GLM tests did control appropriately the false-discovery rate (FDR) at a given scale. We then observed on realistic simulations that there was no substantial inflation of the FDR across scales, as long as the FDR was controlled independently within each scale, and the presence of true associations could be established using an omnibus permutation test combining all scales. Second, we observed both on simulations and on three real resting-state fMRI datasets (schizophrenia, congenital blindness, motor practice) that the rate of discovery varied markedly as a function of scales, and was relatively higher for low scales, below 25. Despite the differences in discovery rate, the statistical maps derived at different scales were generally very consistent in the three real datasets. Some seeds still showed effects better observed around 50, illustrating the potential benefits of multiscale analysis. On real data, the statistical maps agreed well with the existing literature. Overall, our results support that the multiscale GLM connectome analysis with FDR is statistically valid and can capture biologically meaningful effects in a variety of experimental conditions.
\end{abstract}

\begin{keyword}
fmri \sep general linear model \sep functional brain parcellation \sep multiple comparison \sep false discovery rate \sep multiscale analysis \sep connectome
\end{keyword}
\end{frontmatter}


\section*{Highlights}

\begin{itemize}
 \item A mass-univariate GLM analysis on connectomes was estimated multiple times using functional brain parcellations at different scales.
 \item A new omnibus test pooling GLM discoveries across all scales.
 \item On simulations, the FDR was controlled within scale and, in the presence of a significant omnibus test, no marked inflation of the FDR was observed across scales. 
 \item On three real datasets, the statistical association maps generated at different scales were mostly consistent.
 \item In all experiments, the highest discovery rates were found below scale 25, yet some effects were better seen around scale 50. 
\end{itemize}

\section{Introduction}

\paragraph{Context} 
Brain connectivity in resting-state fMRI has been found to be associated with a wide variety of clinical disorders \citep{Fox2010,Castellanos2013,Barkhof2014}. Rather than focusing on a limited set of \emph{a priori} regions of interest, a recent trend is to perform statistical tests of association across the whole connectome, i.e. at every possible brain connection \citep{Shehzad2014}. Such connectome-wide association studies (CWAS) critically depend on the choice of the brain parcels that are used to estimate the connections. Analysis have been performed at different \emph{scales} in the literature \citep{Meskaldji2013}, e.g. voxels \citep{Shehzad2014}, regions \citep{Wang2007a}, or distributed networks \citep{Jafri2008,Marrelec2008}. The main objective of this work was to develop a statistical framework to study the impact of the spatial scale on the results of a CWAS.

\paragraph{Mass-univariate connectome-wide association studies}
The mass-univariate approach to CWAS \citep{Worsley1998} consists of independently estimating a GLM at every connection. In the GLM, a series of equations are solved to find a linear mixture of explanatory variables (called covariates) that best fit the connectivity values observed across the many subjects. A $p$ value is generated for each connection to quantify the probability that the estimated strength of association between this connection and a covariate of interest could have arisen randomly in the absence of a true association \citep{Worsley1995}. The significance level of each test needs to be corrected for the total number of tests, i.e. the number of brain connections, e.g. using random field theory \citep{Worsley1998} or FDR \citep{Benjamini1995}. Correction for multiple comparisons however generally comes at the cost of a sharp decrease in sensitivity.

\paragraph{Multiscale parcellations and testing} 
A straightforward way to mitigate the impact of multiple comparisons on statistical power is to reduce the number of brain parcels. For example, the AAL template \citep{Tzourio-Mazoyer2002} includes 116 brain parcels based on anatomical landmarks. Data-driven algortihms can also generate functional brain parcels \citep{Bellec2006a,Thirion2006a,Craddock2012,Blumensath2013,Thirion2014,Gordon2014}. Few investigators have examined how scale impacts the results of a GLM-connectome analysis. \cite{AbouElseoud2011} explored the impact of the number of components in a dual-regression independent component analysis on the difference between patients suffering from non-medicated seasonal affective disorder and normal healthy controls. The authors concluded that the number of significant findings was maximized at scale 45 (in this case, 45 independent components). The impact of the number of brain parcels was also investigated using spatially-constrained spectral clustering \citep{Craddock2012} at much higher scales (from 50 to 3000+) by \cite{Shehzad2014}. The authors concluded that the association between resting-state connectivity and intelligence quotient was consistent across scales. It should be noted that, in the above-mentioned studies \citep{AbouElseoud2011,Shehzad2014}, the authors did not investigate the implications that the replication of statistical tests at multiple scales may have in terms of the control of false positives. We are thus not currently aware of a valid statistical framework to examine the results of a CWAS with data-driven brain parcellations at multiple scales.

\paragraph{Main objectives}
In this paper, we developped a new method to explore multiscale statistical parametric connectome (MSPC). Multiscale functional brain parcellations and associated connectomes were first generated using a group cluster analysis \citep{Bellec2010c,Bellec2013}. A GLM was then applied on the connectomes and the FDR was controlled at each scale independently \citep{Benjamini1995}. We developed an omnibus test to assess the significance of the number discoveries, pooled across all scales. Our first objective was to empirically assess if the FDR was well controlled within scale, and, when the omnibus test was rejected, across scales as well. Our second objective was to evaluate how the scale impacted the results of a MSPC in terms of the number and nature of the discoveries. In particular, we wanted to check if different scales would bring complementary and biologically plausible insights in the results of the GLM. We conducted a series of experiments involving both simulated and real datasets to address these two objectives, which have been summarized, along with the main findings, in Table \ref{tab_summary}.

\paragraph{Simulations}
We generated several simulation datasets. A first set of experiments involved independent tests on purely synthetic data. A second set of experiments were based on a mixture of real fMRI data with synthetic signal, in order to introduce realistic dependencies between tests, both within and across scales. We simulated group differences using a variety of scenarios covering different combinations of effect and sample sizes as well as different proportions of true non-null hypothesis. As part of the simulation study, we also applied MSPC on real fMRI datasets (the Cambridge sample) using random group differences, thus providing insights in the behaviour of the method under the global null hypothesis, where there is no true association to find. 

\paragraph{Real datasets}
We evaluated the MSPC on three real datasets: (1) a study comparing patients suffering from schizophrenia with healthy control subjects (referred to here as the SCHIZO study, with a large sample size, $n=146$); (2) a study on patients suffering of congenital blindness, compared to sighted controls (referred to here as the BLIND study, with a small sample size, $n=31$), and; (3) a study of motor learning, where resting-state data connectivity was compared before and after learning of a motor task (referred to here as the MOTOR study, with a moderate sample size, $n=54$). These three datasets were chosen to assess how the scale impacted the results of the MSPC with a variety of effect and sample sizes. We had strong a priori hypotheses for those three analyses: changes in the visual network for the BLIND dataset, in the motor network for the MOTOR dataset and a more general dysconnectivity in the SCHIZO dataset. These a priori were useful to assess the biological plausibility of the findings. We checked the assumptions of the parametric GLM estimation, to further validate the specificity of the MSPC approach. We finally quantified similarities and differences of the MSPC findings across scales. 

\begin{landscape}

\begin{table}[!ht]
 \begin{center}
  \begin{tabular}{p{6cm}p{6cm}p{10cm}}
   Specific objectives & Experiment(s) & Finding(s)\\
   \hline
   \\Check the assumptions of the parametric GLM tests and the FDR-BH algorithm, within scales.& Assumptions of the GLM were tested on four real data samples at high scales (300+). Simulations of group differences at multiple scales with tests featuring realistic dependencies. & Although trends of departure from normality and homoscedasticity were observed, no tests passed significance (Section \ref{sec_res_real}). In the simulations with dependent tests, potential departures from the assumptions of the GLM of FDR-BH did not compromise the specificity of the FDR-BH procedure (Figure \ref{fig_fdr_multiscale_300_step_10}).\\
   \\Assess the specificity of MSPC in the absence of signal (``global null''). & Test for differences in average connectivity between random subgroups of the Cambridge sample.  & The FWE under the global null was controlled at nominal level by the multiscale omnibus test (Figure \ref{fig_fdr_multiscale_300_step_10}).\\
   \\Assess the specificity of MSPC within and across scales.&  Multiscale simulation of group differences with independent or dependent tests. & The FDR was controlled at nominal level or below within scale (Figures \ref{fig_multiscale_msteps_schizo}, \ref{fig_fdr_multiscale_300_step_10}). The FDR across scales was controlled in most realistic simulations, and slightly liberal (effective FDR of 0.09 for a nominal level of 0.05) in the worst-case scenario, otherwise featuring sensitivity close to zero at high scales (Figures \ref{fig_fdr_multiscale_300_step_10} and \ref{fig_sens_dep_multiscale_300_step_10_perc0}).\\
   \\Assess the sensitivity of MSPC across scales. & Same as above. & The sensitivity varied substantially across scales. This seemed to reflect at least two phenomena: (1) an intrinsic property of the FDR-BH procedure, which looses sensitivity when the number of multiple comparison increases, before reaching a plateau; (2) effect sizes may change as a function of scale (Figures \ref{fig_sens_multiscale_msteps_schizo}, \ref{fig_sens_dep_multiscale_300_step_10_perc0}, \ref{fig_sens_dep_multiscale_300_step_10_perc30}).\\
   \\Assess the plausibility of the results identified with MSPC on real data. & GLM connectome analyses in three real datasets at multiple scales.&The MSPC identified biologically plausible changes in connectivity in all three analyses (Figures \ref{fig_perc_disc_maps}, \ref{fig_eff_maps}).\\
   \\Assess if different scales can identify complementary effects. & Same as above. & The discovery rate was markedly higher at low scales, below 50 (Figures \ref{fig_perc_disc}, \ref{fig_perc_disc_maps}). Statistical parametric maps were highly consistent across scales (Figure \ref{fig_eff_multiscale}), although some effects associated with specific structures were better seen at high scales (Figures \ref{fig_perc_disc_maps}, \ref{fig_eff_maps}).\\
  \end{tabular}
 \end{center}
 \caption{Summary of the specific objectives, experiments and findings of the paper.}
\label{tab_summary}
\end{table}
\end{landscape}

\section{Statistical testing procedures}
\label{sec_theory}
\subsection{Functional parcellations} The first step to build a connectome is to select a parcellation of the brain, with $R$ parcels. In this work, we used functional brain parcellations, aimed at defining groups of brain regions with homogeneous time series. A number of algorithms have been proposed with additional spatial constraints, to ensure that the resulting parcels are spatially connected \citep{Lu2003,Thirion2006a,Craddock2012}. However, from a pure functional viewpoint, the spatial constraint seems somewhat arbitrary, as functional units in the brain at low resolution encompass distributed networks of brain regions with homotopic regions often being part of a single parcel \citep{DeLuca2006,Damoiseaux2006}. Some works have thus used distributed parcels as the spatial units to measure functional brain connectivity, e.g. \citep{Jafri2008,Marrelec2008}. To generate the functional parcelations, we relied on a recent method called ``Bootstrap Analysis of Stable Clusters'' (BASC), which can identify consistent functional parcels for a group of subjects \citep{Bellec2010c}, using a hierarchical cluster with Ward's criterion both at the individual and the group levels. The functional parcels can be generated at any arbitrary scale (within the range of the fMRI resolution), and we considered only parcels generated at the group level, which were non-overlapping and not necessarily spatially contiguous. 

\subsection{Functional connectome}
\label{sec_connectome}

\begin{figure}[!ht]
\begin{center}
\includegraphics[width=\linewidth]{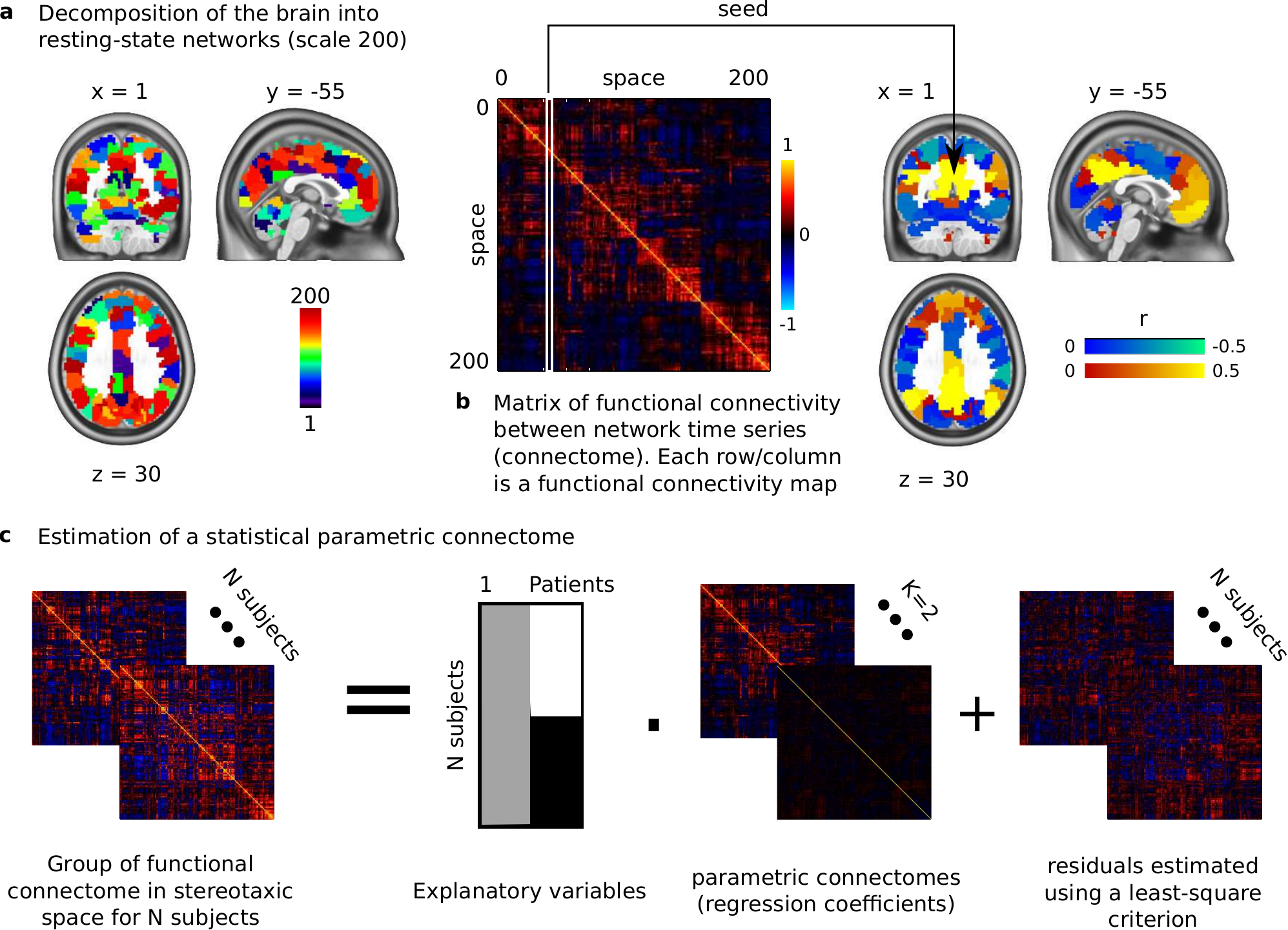}
\end{center}
\caption{
{\bf General linear model applied to connectomes.} {The connectivity is measured between $R$ brain parcels generated through a clustering algorithm (panel \textbf{a}). The connectome is a $R \times R$ matrix measuring functional connectivity between- and within-parcels (panel \textbf{b}). The association between phenotypes and connectomes is tested independently at each connection using a general linear model at the group level (panel \textbf{c}). The results presented here are for illustration purpose only, and not related to the results presented in the application sections of the manuscript.} 
}
\label{fig_glm}
\end{figure}

For each scale, and each pair of distinct parcels $i$ and $j$ at this scale, the between-parcel connectivity $y_{i,j}$ is measured by the Fisher transform of the Pearson's correlation between the average time series of the parcels. Note that other measures can be used to quantify interactions between parcels, such as partial correlations \citep{Marrelec2006a}. We used correlation as it is the simplest, most popular and still fairly accurate \citep{Smith2011} measure of interaction in fMRI. The statistical framework presented here could still be applied to many other measures (see \ref{sec_discussion}). The within-parcel connectivity $y_{i,i}$ is the Fisher transform of the average correlation between time series of every pair of distinct voxels inside parcel $i$. The connectome $\mathbf{Y}=(y_{i,j})_{i,j=1}^R$ is thus a $R\times R$ matrix. Each column $j$ (or row, as the matrix is symmetric) codes for the connectivity between parcel $j$ and all other brain parcels, or in other word is a full brain functional connectivity map. See Figure \ref{fig_glm}\textbf{a-b} for a representation of a parcellation and associated connectome. Connectomes are generated independently at each scale. See \ref{app_connectome} for a more formal description of the connectome generation.

\subsection{Statistical parametric connectome}
\label{sec_spc}
For a scale with $R$ parcels, there are exactly $L=R(R+1)/2$ distinct elements in an individual connectome $\mathbf{Y}$. This connectome can be stored as a $1\times L$ vector, where the brain connections have been ordered arbitrarily along one dimension. When functional data is available on $N$ subjects, the group of connectomes is then assembled into a $N\times L$ array $\mathbf{Y}=(y_{n,l})$, where $n=1,\dots,N$ each code for one subject and $l=1,\dots,L$ each code for one connection. A general linear model (GLM) can then be used to test the association between brain connectivity and a trait of interest, such as the age or sex of participants. All of these $C$ explanatory variables are entered in a $N\times C$ matrix $\mathbf{X}$. The variables are typically corrected to have a zero mean across subjects, and an intercept (i.e. a column filled with 1) is added to $\mathbf{X}$. The GLM relies on the following generative model:
 \begin{equation}
 \label{eq_glm}
  \mathbf{Y} = \mathbf{X}\mathbf{B} + \mathbf{E},
 \end{equation}
\begin{itemize}
 \item $\mathbf{Y}$ is a $N\times L$ matrix where each row codes for a subject, and each column codes for a connection,
 \item $\mathbf{X}$ is a $N \times C$ matrix of explanatory variables (or covariates) where each row codes for a subject and each column codes for a covariate,
 \item $\mathbf{B}$ is an unknown $C\times L$ matrix of linear regression coefficients where each row codes for a covariate and each column codes for a connection,
 \item $\mathbf{E}$ is a $N\times L$ random (noise) variable, with similar coding as $\mathbf{Y}$.
\end{itemize}
We relied on the following parametric assumptions on the noise $\mathbf{E}$ are (1) that its rows are independent; (2) that each element follows a normal distribution with zero mean, and (3) that the variance of all elements are constant within a column, also called the homoscedasticity assumptions. As the data generated from different subjects are statistically independent the first assumption is reasonable. We tested the normality and homoscedasticity assumptions on real datasets. Under these parametric assumptions, the regression coefficients $\mathbf{B}$ can be estimated with ordinary least squares and, for a given ``contrast'' vector $\mathbf{c}$ of size $1\times C$, the significance of $\mathbf{c}\hat{\mathbf{B}}$ can be tested with a connectome of $t$-test $(t_{l})_{l=1}^L$, with associated $p$-values $(p_l)_{l=1}^L$. The quantity $p_l$ controls for the risk of false positive findings at each connection $l$. The GLM applied on connectomes is illustrated in Figure \ref{fig_glm}\textbf{c}. See \ref{app_glm} for the equations related to the estimation and testing of regression coefficients in the GLM.

\subsection{The Benjamini-Hochberg FDR procedure} The number $L$ of tests $(p_l)_l$ grows quadratically with the scale $K$. The significance value applied on $p_l$ within a scale thus needs to be adjusted for this multiple comparison problem. We implemented the benjamini-Hochberg (BH) procedure \citep{Benjamini1995} to control the FDR at a specified level $\alpha$ within scale. The idea of the FDR is not to strictly control the probability to observe at least one false positive (a quantity know as family-wise error, FWE), but rather to control, on average, the proportion of false positive amongst the findings. Note that controlling for the FDR is not necessarily a more liberal attitude than controlling for the FWE: if the global null hypothesis is verified, i.e. all discoveries are false positive, then the FDR is exactly the FWE. In the presence of true discoveries, however, the FDR procedure tolerates in general more noise than a FWE approach. The actual number of false discoveries will depend on the amount of signal (true positive) present in the data, and is therefore a context-dependent question. The BH procedure is a popular technique to control the FDR, which was designed for independent tests. The BH still has been shown to have a satisfactory behaviour even in the presence of positive correlation between the tests $p_l$ \citep{Benjamini2001}. On simulations, the specificity of the FDR-BH algorithm was assessed in the presence of a realistic amount of correlation between tests, as would be found in a MSPC analysis of fMRI data. 

\subsection{Multiscale parcellations}

\begin{figure}[!ht]
\begin{center}
\includegraphics[width=\linewidth]{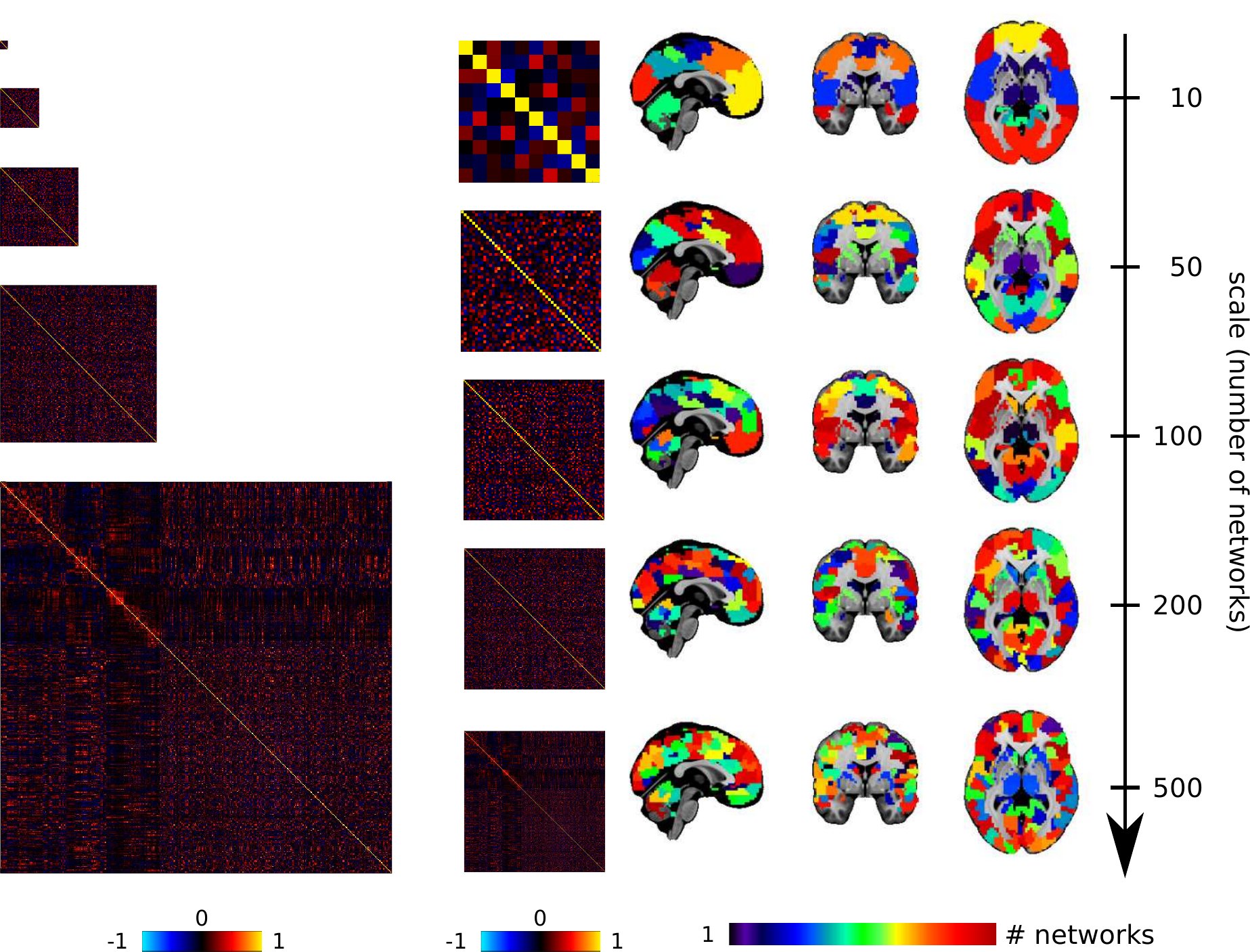}
\end{center}
\caption{
{\bf General linear model applied to connectomes at multiple spatial scales.} {The generation of data-driven brain parcels is iterated at different scales (number of parcels), using the bootstrap analysis of stable clustered (BASC), with a hierarchical clustering using Ward's criterion. The statistical parametric connectomes are represented using both their real size (left column) and after rescaling to fit identical size (middle column) to illustrate the quadratic increase in the number of connections (multiple comparisons) that comes with an increase in the number of parcels. The results presented here are for illustration purpose only, and not related to the results presented in the application sections of the manuscript.} 
}
\label{fig_connectome_multiscale}
\end{figure}

To explore the association of connectomes and a trait of interest at multiple scales, some multiscale brain parcellations first need to be generated, for which we employed the BASC method \citep{Bellec2010c}. A GLM was then estimated with FDR control at each scale independently (Figure \ref{fig_connectome_multiscale}). To choose the scales of analysis, a comprehensive strategy would consist of a regular grid, e.g. from $10$ to $300$ brain parcels, with a step of $10$. The GLM results at such scales may however be highly redundant, as some parcels may be found identically at different scales, if those are close. To systematically examine the results of the GLM analysis at all selected scales, a better strategy would be to select a limited number of non-redundant scales that span a given range (e.g. $10$ to $300$). For this purpose, we used the multiscale stepwise selection (MSTEPS) algorithm recently proposed by \cite{Bellec2013} to select scales that provide an accurate summary of the stable features of brain clusters observed at all possible scales. Both strategies (comprehensive grid of scales and sparse subset) were examined in this work, both on simulations and real data.

\subsection{Multiscale testing}
Testing GLM on connectomes at multiple scales introduces a new level of multiple comparisons, this time across scales rather than across connections. In addition to the FDR within scale, one now needs to consider the FDR across scales, i.e. for all tests combined across scales. For example, if two scales were used, $K=10$ and $K=100$, there would be $K(K+1)/2$ tests at each scale (i.e. $55$ and $5050$). If there were $10$ discoveries at scale $10$, $1$ of which was a false positive, and $200$ discoveries at scale $100$, $10$ of which were false positive, the ratio of false discoveries for this simulation would be $1/10=0.1$ within scale 10, and $10/200=0.05$ within scale $100$. The ratio of false discoveries across scales would be $(1+10)/(10+200)=0.052$. The FDR within-scale and between-scale would be the average of these false discovery proportions over many replications of such experiments. A straightforward way to control the FDR across scales would be to apply the FDR-BH procedure to the whole set of tests, pooled across scales. For a regular grid of scales ranging from 10 to 300, with a step of 10, the largest numbers of tests within scale is $300\times 301/2=45150$. The total number of tests is in that case $475075$, an order of magnitude larger. Because the FDR-BH relies on a Bonferroni estimation of the number of false positives, an FDR procedure in such a high dimension is bound to have low sensitivity, and would be a high price to pay to explore multiple scales of a GLM analysis. However, we can note that in a multiscale analysis, a detection at any scale makes the presence of signal at all the other scales very likely. In the presence of multiple families of tests and a substantial amount of true discoveries, \citep{Efron2008} hypothesized based on simulations that there would be no need to adjust the FDR across families, when the FDR is controlled within family. The rationale for this hypothesis is that, in the presence of signal, the FDR controls for a proportion, which behaves well when multiple families are combined. By contrast, the FWE looks at a maximum statistics and gets mechanically inflated when mutliple families are combined. In the context of a MSPC analysis, Efron's hypothesis would mean that controlling for the FDR at each scale independently would also guarantee that the FDR would be controlled across scales. If supported empirically, this hypothesis would justify to examine the results of a GLM on connectomes at many scales, without introducing the need for any additional correction of multiple comparisons beyond what is required by a traditional, single-scale analysis. The multiscale exploration would basically come for free, provided that enough true discoveries were made overall. 

\subsection{Omnibus test}

Our hypothesis is that the control of the overall FDR (across connections and scales) can be attained through the sole control of FDR within scale, in the presence of a substantial volume of true discoveries. At a given scale $s$, $V_s$ is the percentage of discoveries, i.e. the number of significant tests as identified by FDR-BH at a given level $\alpha$, divided by the total number of tests. For a grid of scales, e.g. those selected by MSTEPS, the overal volume of discoveries $V$ was defined as the average of $V_s$ across all scales $s$. As a minimal requirement for the volume of true discoveries to be ``substantial'', we developped an omnibus test to reject the hypothesis that $V$ could be observed under the global null hypothesis $(\mathcal{G}_0)$, i.e. \emph{no non-null effect at any connection and any scale}. This test proceeded by comparing the volume of discoveries $V$ observed empirically in the group sample against the volume of discoveries $V^*$ that could be observed under $(\mathcal{G}_0)$. To generate replications of $V^*$ under $(\mathcal{G}_0)$, the GLM was first applied with a reduced model where the explanatory variable of interest (as selected through the contrast) had been removed. Then, a permutation of the residuals was generated as described in \citep{Anderson2002}, see \ref{app_omnibus}. A replication of connectomes was generated under $(\mathcal{G}_0)$ by adding the permuted residuals to the estimated mixture of reduced explanatory variables. In order to respect the dependencies between connectivity estimates within and across scales, the same permutation of the subjects was applied to all of the connections and scales. The detection procedure was applied on each replication of connectomes, under $(\mathcal{G}_0)$, and the total volume of discoveries was derived. A Monte-Carlo approximation, with typically $10000$ permutation samples on real data, was used to estimate a false-positive rate $p$ when testing against the global null hypothesis. Note that a single omnibus test was derived, controlling for the FWE of the experiment as a whole. If this test passed significance, each scale was examined with a control of the FDR at $\alpha=0.05$, uncorrected for multiple comparisons across scales. If the omnibus test did not reach significance, then no connection at any scale was deemed significant.
\section{Evaluation on simulated datasets with independent tests}
\label{sec_simus_ind}

\subsection{Methods}

\paragraph{Data-generating procedure} We started by simple simulations of independent tests, to assess to which extent the hypothesis of \cite{Efron2008} was robust to different scenarios, and if the omnibus test would systematically ensure that the FDR across scales would be well controlled. A number $K$ of test families were generated independently, each one composed of $L_k$ tests, $k=1,\dots,K$. Each family included a set proportion of true non-null hypotheses $\pi_1$, identical for all families. If  $\pi_1 L_k$ was not an integer, the number of true positives $n_k$ was set to either $\lfloor \pi_1 L_k \rfloor$ or $\lfloor \pi_1 L_k \rfloor +1$, with probabilities such that on average over many simulations $\mathbb{E}(n_1)=\pi_1 L_k$. For a non-null test $l$, the associated $p$-value was simulated as:
\begin{equation}
 y_l = \theta+z_l, z_l\sim \mathcal{N}(0,1),
\end{equation}
\begin{equation}
 p_l = \textrm{Pr}\left(x \geq y_l| x \sim \mathcal{N}(0,1)\right),
\end{equation}
where $\mathcal{N}(0,1)$ was a Gaussian distribution with zero mean and unit variance, and $\theta>0$ was a simulation parameter (further called effect size). The null tests were generated the same way, but with an effect size $\theta=0$. 

\paragraph{Simulations scenarios} For each experiment, all combinations of effect size in the grid $\{2,3,5\}$ and $\pi_1$ in the grid $\{0\%,1\%,2\%,5\%,10\%\}$ were considered. We implemented a series of experiments:
\begin{itemize}
 \item We first checked how the FDR across scales behaved as a function of the number of families $K$, with $K$ in $\{2,5,10\}$ and $L_k$ equals to $1000$ (corresponding approximately to the number of tests at scale 45).
 \item We then checked how the FDR across scales behaved as a function of the number of tests per family $L_k$ with $L_k$ identical for all $k$, and in the grid $\{100,1000,10000\}$ (corresponding roughly to scales $14$, $45$ and $141$), and $K=5$ families. 
 \item We checked how the FDR across families behaved for a number of families and a number of tests per family that would be comparable to situations encountered on multiscale GLM-connectome analysis. 
 \begin{itemize}
 \item We first tested the scales selected by MSTEPS on the SCHIZO dataset (see Section \ref{sec_real}), i.e. $K=7$ and the $(L_k)_k$ equal to ($28$, $136$, $325$, $1540$, $6555$, $19900$, $53956$), corresponding to the number of tests at scales $7$, $16$, $25$, $55$, $114$, $199$, $328$. 
 \item We then tested the procedure on $K=30$ and $(L_k)_k$ ranging from $55$ to $45150$, which would be equal to the number of tests associated with a regular grid covering scales $10$ to $300$ with a step of $10$. 
 \item We finally tested the behaviour of smaller grids, with a number of tests equivalent to GLM tests over scales ranging from $10$ to either $50$, $100$ or $300$ (with a step of 10).
 \end{itemize}
\end{itemize}

\paragraph{Computational environment} All the experiments reported in the paper were performed using the NeuroImaging Analysis Kit (NIAK\footnote{\url{http://www.nitrc.org/projects/niak/}}) version 0.12.18, under CentOS version 6.3 with Octave\footnote{\url{http://gnu.octave.org}} version 3.8.1 and the Minc toolkit\footnote{\url{http://www.bic.mni.mcgill.ca/ServicesSoftware/ServicesSoftwareMincToolKit}} version 0.3.18. Analyses were executed in parallel on the "Guillimin" supercomputer\footnote{\url{http://www.calculquebec.ca/en/resources/compute-servers/guillimin}}, using the pipeline system for Octave and Matlab \citep{Bellec2012}, version 1.0.2. The scripts used for processing can be found on Github\footnote{\url{https://github.com/SIMEXP/glm_connectome}}.

\paragraph{Statistical testing procedure} For each simulation scenario, the FDR-BH procedure was applied to each family independently, with a significance level $\alpha$ in the grid $\{0.01,0.05,0.1,0.2\}$. To estimate the distribution of the volume of discovery under the global null, 1000 samples were generated with the parameters $\theta$ and $\pi_1$ set to zero, for each choice of $K$ and $(L_k)_k$. These replications under the globall null were used to generate the $p$-values of the omnibus test for all simulations with identical $K$ and $(L_k)_k$.

\paragraph{Effective FDR, sensitivity and omnibus test} The effective FDR at each scale was computed as the number of false discoveries divided by the total number of discoveries, averaged across 1000 simulation replications for each simulation scenario. The effective sensitivity was computed at each scale as the number of true discoveries, divided by the number of true non-null hypotheses present at this scale, and averaged across the 1000 replications. To compute the FDR across scales, the same procedure to estimate the effective FDR was applied to the combination of tests pooled across all families. Finally, we also derived a modified FDR and sensitivity, where the BH-FDR procedure was combined with the omnibus permutation test, at a significance level of $p<0.05$. 

\subsection{Results}

\begin{figure}[!ht]
\begin{center}
\includegraphics[width=\linewidth]{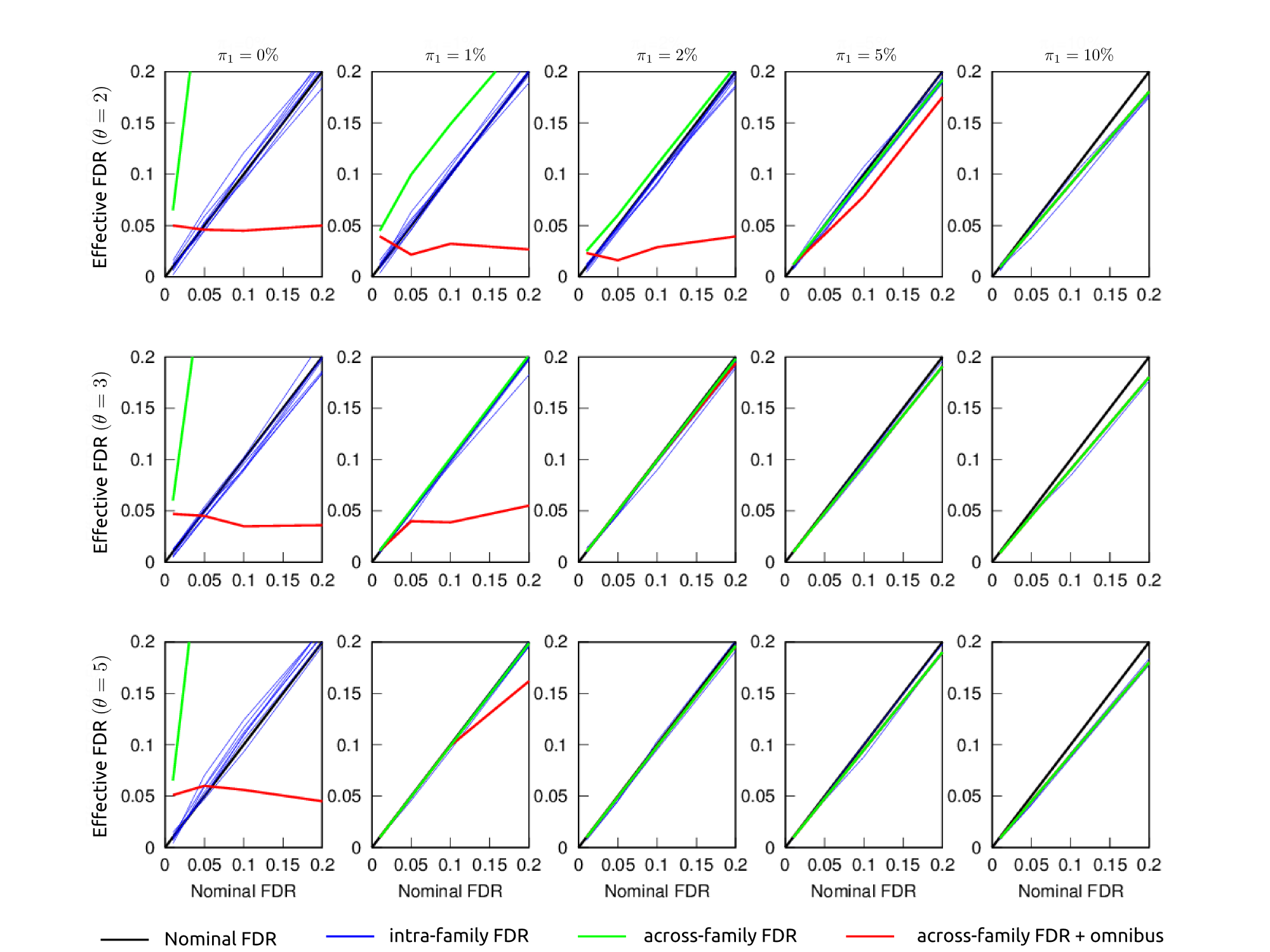}
\end{center}
\caption{
{\bf Nominal vs effective FDR on simulations with independent tests ($K=7$, $L_k$ in $(28,136,325,1540,6555,19900,53956)$, corresponding to the number of tests associated with the scales selected by MSTEPS on the SCHIZO dataset.} {The effective FDR is plotted against the nominal FDR within each family (blue plots), across all families (green plots) and across all families, combined with an omnibus test for rejection of the global null hypothesis (red plot). The expected (nominal) values are represented in black plots, corresponding to the four tested FDR levels: $0.01, 0.05, 0.1 ,0.2$. Each column corresponds to a certain proportion of non-null hypothesis per family $\pi_1$ ($0\%,1\%,2\%,5\%,10\%$), and each row corresponds to a different effect size $\theta$ ($2,3,5$), see text for details. Please note that in the presence of strong signal (large $\theta$ and/or $\pi_1$), the omnibus test is always rejected, and the green plot matches perfectly the red plot, which becomes invisible.} 
}
\label{fig_multiscale_msteps_schizo}
\end{figure}

\paragraph{The intra-family FDR} Some of the conclusions in that example were observed consistently across all our experiments, see for instance Figure \ref{fig_multiscale_msteps_schizo}. In particular, the effective FDR within each family is following closely $(1-\pi_1)\alpha$. This is a replication of a well-established result: the BH-FDR procedure is conservative by a factor $(1-\pi_1)$ for independent tests. For example, for $\alpha=0.2$ and a $\pi_1$ of $10\%$, the effective FDR is of approximately 0.18 for all effect sizes $\theta$, see the right-most values in the last column in Figure \ref{fig_multiscale_msteps_schizo}. 

\paragraph{The FDR across families} The FDR across families followed a smooth transition between two regimes. The first regime, called ``liberal'' was clearly seen under the global null hypothesis (first column in Figure \ref{fig_multiscale_msteps_schizo}). In this regime, the FDR matched with the FWE, i.e. the probability to have one or more false positive. As expected, controling the FWE within family did not control for the FWE across families, i.e. the effective FDR across families was largely superior to the prescribed level $\alpha$. The second regime, called ``exact'' was seen best in the last column of Figure \ref{fig_multiscale_msteps_schizo}. In the presence of a large $\pi_1$ (in this example $10\%$), the FDR across scales precisely followed the FDR within scales. The transition between these two regimes (liberal and exact) was smooth, and in situations that ressembled the global null hypothesis (i.e. at low $\pi_1$ or effect size), the FDR across scales was more liberal than the nominal $\alpha$, sometimes by a wide margin, e.g. $L_k=1000$, $\pi_1=1\%$ and $\theta=2$ in Figure \ref{fig_fdr_variable_K}. Note that both increasing the effect size, or increasing $\pi_1$ both pushed the FDR across families towards the ``exact'' regime. 

\paragraph{The FDR across families, with omnibus test} The effect of the omnibus test on the FDR across families was particularly apparent under the global null hypothesis in all simulations: the FDR across scales matched the FWE, which was less than, or equal to the $p$ value of the omnibus test, as expected, see the first column in Figure \ref{fig_multiscale_msteps_schizo}. More generally, for simulation scenarios that represented a transition between the liberal and exact regimes of the FDR across families, the application of the omnibus test tended to make the FDR across families more conservative. There was still no guarantee that the FDR across families conformed to the specified $\alpha$ level, as can be seen for $L_k=1000$, $\pi_1=1\%$ and $\theta=2$ in Supplementary Figure \ref{fig_fdr_variable_K}, where the effective FDR was larger than $0.1$ for a nominal FDR of $0.05$. This behaviour was however found to be very much dependent on the number of tests per family and the number of families, which was further studied below.

\paragraph{Influence of the number of families $K$} By varying $K$ in $\{2,5,10\}$ for a fixed number of tests per family ($L_k=1000$ for all $k$), we found that the transition between the liberal and exact regime of the FDR across scales took longer when the number of families increased. For $\pi_1=2\%$ and $\theta=2$, and a nominal FDR $\alpha=0.05$, the effective FDR was about $0.07$ with $K=2$, while it increased to almost $0.1$ for $K=10$ (Supplementary Figure \ref{fig_fdr_variable_K}).

\paragraph{Influence of the number of tests per family $L$} By varying $L_k$ in $\{100,1000,10000\}$ (with $L_k$ identical for all $k$) for a fixed number of families $K=5$, we found that the transition between the liberal and exact regime of the FDR was quicker when the number of tests per family increased. In other words, the exact regime appears as an asymptotic behaviour of the FDR across scales, when the number of tests per family becomes large. For example, for $\pi_1=2\%$ and $\theta=2$, and a nominal FDR $\alpha=0.05$, the effective FDR across scales went from above $0.1$ with $100$ tests per family to below $0.06$ with $10000$ tests per family (Supplementary Figure \ref{fig_fdr_variable_L}). 

\begin{figure}[!ht]
\begin{center}
\includegraphics[width=\linewidth]{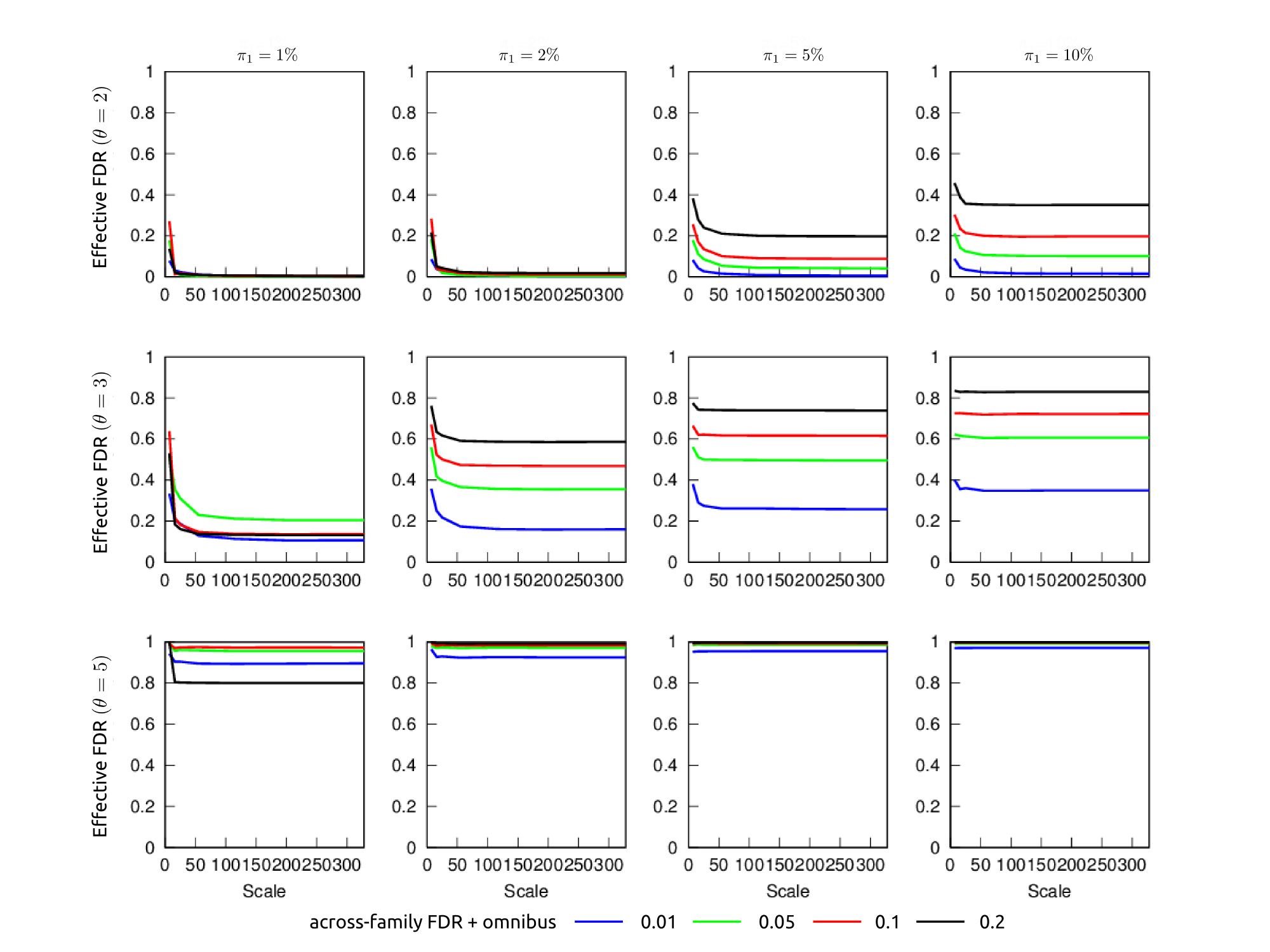}
\end{center}
\caption{
{\bf Sensitivity on simulations with independent tests, $K=7$, $L_k$ in $(28,136,325,1540,6555,19900,53956)$, corresponding to the number of connections associated with the scales selected by MSTEPS on the SCHIZO dataset.} {The sensitivity is plotted as a function of scales at four tested (within-scale) FDR levels: $0.01, 0.05, 0.1 ,0.2$. A test is only considered as significant if in addition an omnibus test against the global null hypothesis across scales as been rejected at $p<0.05$. Each column corresponds to a certain proportion of non-null hypothesis per family $\pi_1$ ($0\%,1\%,2\%,5\%,10\%$), and each row corresponds to a different effect size $\theta$ ($2,3,5$), see text for details. } 
}
\label{fig_sens_multiscale_msteps_schizo}
\end{figure}

\paragraph{Scenarios based on multiscale connectome testing} Figure \ref{fig_multiscale_msteps_schizo} shows the effective FDR for $K=7$ families including a number of tests similar to the ones associated with the scales selected by MSTEPS on the SCHIZO dataset. In this setting, and despite the high number of families, the transition from the liberal to the exact regime was very quick, which likely reflected the high number of tests found in most families. In particular, the FDR was found to be appropriately controlled across scales as soon as $\pi_1>5\%$. In situations were the FDR across scales was liberal, the correction by the omnibus tests made the test very close to nominal levels, or conservative. Regarding the sensitivity of the tests, as expected, increasing either $\pi_1$ or $\theta$ increased the overall sensitivity of the tests. We observed a general trend in all scenarios: the sensitivity peaked at very low scales, and decreased exponentially to reach a plateau around scale 10 to 50. After this initial loss in sensitivity, the sensitivity was uniform across scales (Figure \ref{fig_sens_multiscale_msteps_schizo}). Identical conclusions were reached with families akin to connectome testing on a regular grid of scales ranging from 10 to either 50, 100, or 300 parcels (with a step of 10). See Supplementary Figure \ref{fig_multiscale_variable_grid} for the effective FDR, and Supplementary Figure \ref{fig_sens_multiscale_300_step_10} for sensitivity results. 

\section{Evaluation on simulated datasets with dependent tests}
\label{sec_simus_dep}
\subsection{Methods}
\label{sec_methods_simu}

\paragraph{Data-generating procedure} We designed a simulation framework for multiscale GLM-connectome analysis in the presence of depencies between tests, both within scale and across scales. To ensure that these depencies would be as realistic as possible, semi-synthetic datasets were generated starting from a large real sample (Cambridge) released as part of the 1000 functional connectome project\footnote{\url{http://fcon_1000.projects.nitrc.org/fcpClassic/FcpTable.html}} \citep{Biswal2010}. This sample \citep{Liu2009} included resting-state fMRI time series (eyes opened, TR of 3 seconds, 119 volumes per subject) collected with a 3T scanner on 198 healthy subjects (75 males), with an age ranging from 18 to 30 years. All the datasets were preprocessed and resampled in stereotaxic space, as described in Section \ref{sec_real_methods}. A region growing algorithm was used to extract 483 regions, common to all subjects, as described in \cite{Bellec2010c}. For each subject, the average functional time series and associated connectomes were generated using these regions\footnote{The average time series have been publicly released at \url{http://figshare.com/articles/Cambridge_resting_state_fMRI_time_series_preprocessed_with_NIAK_0_12_4/1159331}.} (see Section \ref{sec_connectome}). The average connectome across all subjects was derived, and a hierarchical clustering procedure (with Ward's criterion) was applied to derive a hierarchy of brain parcels at all possible scales, ranging from 1 to 483. The simulation procedure relied on the manual selection of a critical scale $K$ and a particular cluster $k$ at this scale. For each simulation, two non-overlapping subgroups of subject ($N$ subjects per group) were randomly selected. A circular block bootstrap (CBB) procedure was applied to resample the individual time series, using identical time blocks within each cluster, and independent time blocks in different clusters. This resampling scheme ensured that within-cluster correlations were preserved, while between-cluster correlations had a value of zero on average. Finally, for the subjects selected to be in the first group, a single realization of a independent and identically distributed Gaussian variable, where each time point had a zero mean and a variance of $a^2$, was added to the time series of the regions inside cluster $k$, after the time series were corrected to a zero temporal mean and a variance of $(1-a^2)$. The addition of this signal increased the intra-parcel connectivity of the 
cluster including cluster $k$ for all scales smaller or equal to $K$, and increased the within- as well as between-parcel connectivity for all clusters included in cluster $k$ for scales strictly larger than $K$. Because of the absence of correlations between parcels at scale $K$ (due to the CBB resampling), all other connections within- or between clusters at every scale were left unchanged by this procedure. It was thus possible to know exactly which connections were true or false null hypothesis in the group difference at every scale. Supplementary Figures \ref{fig_generation_simu} and \ref{fig_generation_simu_multiscale} outline the procedure of multiscale connectome simulation.

\paragraph{Effect size and proportion of non-null hypothesis} A number of clusters of reference were handpicked such that the proportion of non-null hypothesis $\pi_1(k)$ would be about $1\%$, $2\%$, $5\%$ and $10\%$ at all scales $k$. Note that these reference clusters were used to set true non-null hypotheses at all the scales of analysis, yet the subdivisions (or merging) associated with these clusters represented a varying proportion of the number of clusters at any given scale. As a consequence, and unlike simulations of independent tests, $\pi_1(k)$ was dependent on the scale $k$. Two values for $a^2$ were selected: $0.1$ and $0.2$. The effect size associated with a given $a^2$ actually depended on the within-cluster correlations, between-subject variance in connectivity as well as the scale of analysis. Two sample sizes were investigated: $N=40$ ($20$ subject per group), and $N=100$ ($50$ subjects per group). See Supplementary Material Figure \ref{fig_eff_size_dep_multiscale_300_step_10} for plots of the effect size and percentage of true non-null hypotheses as a function of scale. For each simulation scenarios, $1000$ Monte-Carlo samples were generated and subjected to the MSPC analysis, with FDR levels of $0.01$, $0.05$, $0.1$ and $0.2$, as well as an omnibus test at $p<0.05$. The same scales were tested here as in the simulations for independent tests: the scales selected by MSTEPS on the SCHIZO experiment, and a regular grid from $10$ to $300$ clusters (with a step of 10).

\paragraph{Simulations under the global null} To assess the behaviour of the testing procedures in the absence of any signal, we also ran experiments under the global null. In that case real connectomes were generated for randomly selected and non-overlapping groups of subjects, and then a testing procedure was implemented to assess the significance of group differences. In these experiments, no bootstrap was performed on individual time series nor any signal was added. The experiments simply consisted in comparing real connectomes between random groups of subjects sampled from identical populations, using real dependencies between tests.

\paragraph{Robustness to the choice of clusters} Finally, we also investigated how the procedure behaved when the clusters used in the testing procedures did not match exactly with the clusters that were used to generate the simulations. For this purpose, for each simulation, no signal was actually generated in 30\% of the regions in the cluster of reference, but rather in an equivalent number of arbitrarily selected regions from other clusters. The same regions were selected across all simulations to simulate a systematic departure of the test clusters from the ground truth clusters. The multiscale clusters without perturbations were used in the statistical testing procedures. In this setting, many connections outside of the cluster of reference ended up with very small effects, and we did not investigate the specificity given the very large number of true non-null hypotheses and large variations in effect size. However, we did investigate the sensitivity of the FDR-BH procedure, using the same definition of true non-null hypothesis as with the simulations without perturbation. 

\subsection{Results}

\begin{figure}[!ht]
\begin{center}
\includegraphics[width=\linewidth]{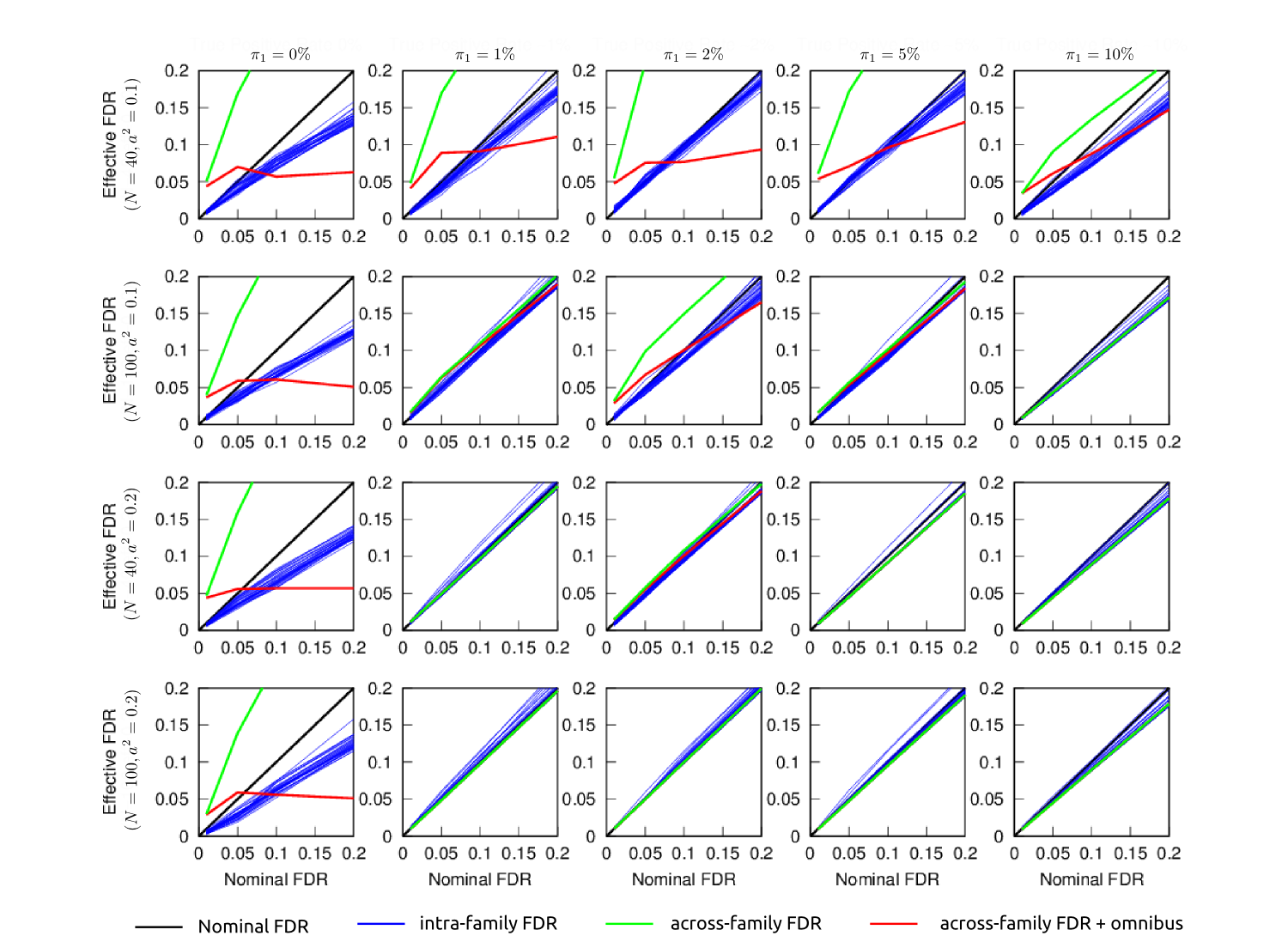}
\end{center}
\caption{
{\bf Nominal vs effective FDR on simulations with dependent tests ($K=30$, $L_k$ ranging from $55$ to $45150$, corresponding to the number of connections associated with a regular grid of scales covering 10 to 300 with a step of 10).} {The effective FDR is plotted against the nominal FDR within each family (blue plots), across all families (green plots) and across all families, combined with an omnibus test for rejection of the global null hypothesis (red plot). The expected (nominal) values are represented in black plots, corresponding to the four tested FDR levels: $0.01, 0.05, 0.1 ,0.2$. Each column corresponds to a certain proportion of non-null hypothesis per family $\pi_1$ ($0\%,1\%,2\%,5\%,10\%$), and each row corresponds to a different combination of effect and sample size $N$ in $\{40,100\}$, $a^2$ in $\{0.1,0.2\}$, see text for details. Please note that in the presence of strong signal (large $\theta$ and/or $\pi_1$), the omnibus test is always rejected, and the green plot matches perfectly the red plot, which becomes invisible.} 
}
\label{fig_fdr_multiscale_300_step_10}
\end{figure}

\paragraph{Effective FDR within scale} Figure \ref{fig_fdr_multiscale_300_step_10} represents the empirical FDR as a function of scale for the MSPC procedure, in the case of a regular grid of scales covering $10$ to $300$ brain parcels and a perfect match between the true and test clusters. The effective FDR was conservative within scale on the simulations with dependent tests, e.g. the effective FDR was about $0.15$ for a nominal FDR of $0.2$. This is in contrast with the independent tests, where the control of the FDR within scale was exact under the global null hypothesis. Our interpretation was that the large positive correlations present in fMRI time series caused the FDR-BH procedure to become conservative. In the presence of signal, the FDR within scale was still well controlled, with the same $(1-\pi_1)$ factor on the effective FDR as was observed with independent tests. 

\paragraph{Effective FDR across scales} As was observed on independent tests, the FDR across scales transitioned between a ``liberal'' regime, in simulation scenarios close to the global null hypothesis, to an exact regime, where the FDR across scales matched the FDR within scales. The transition between regimes happened quite fast, with either $a^2=0.2$ or $N=100$, as soon as $\pi_1$ was larger than $5\%$. When combined with the omnibus test at $p<0.05$, the FWE under the global null hypothesis became exact or conservative for a FDR level above $0.05$. Importantly, the omnibus test also made the procedure either conservative (for $\alpha\geq 0.1$) or only slightly liberal in the scenarios were the FDR across scales transitioned between the ``liberal'' and ``exact'' regimes, with the effective FDR in the range $0.06$ to $0.09$ for a nominal level of $0.05$ in the worst cases (i.e. $N=40$, $a^2=0.1$ and $\pi_1=1\%$). The conclusions were identical when using a regular grid of $K=30$ scales ranging from 10 to 300 parcels (with a step of 10), or $K=7$ scales identical to those selected by MSTEPS on the SCHIZO dataset, see Supplementary Figure \ref{fig_simu_dep_msteps_schizo}. 

\begin{figure}[!ht]
\begin{center}
\includegraphics[width=\linewidth]{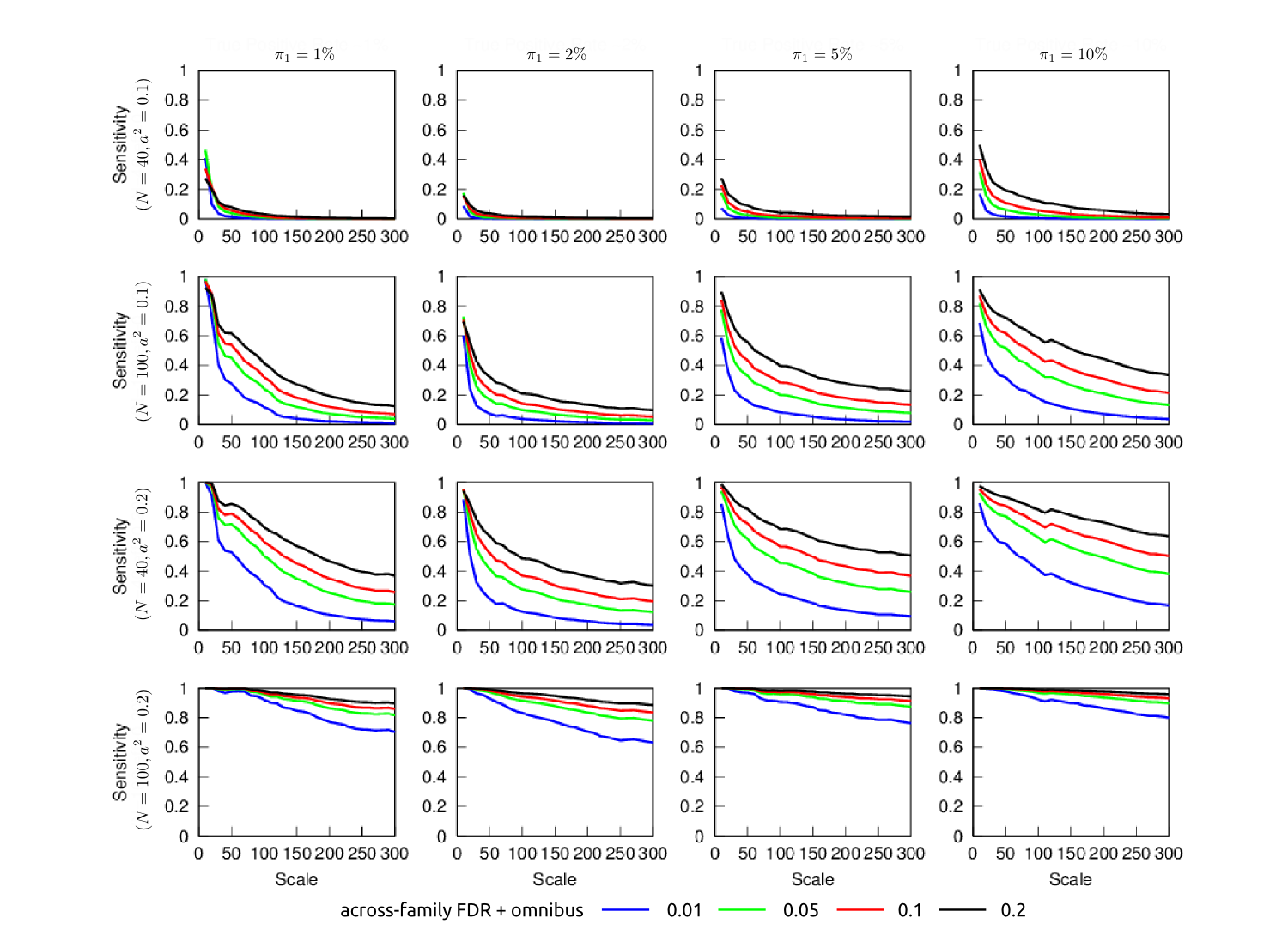}
\end{center}
\caption{
{\bf Sensitivity on simulations with dependent tests, when no mismatch is introduced between the true and test clusters ($K=30$, $L_k$ ranging from $55$ to $45150$, corresponding to the number of connections associated with a regular grid of scales covering 10 to 300 with a step of 10).} {The sensitivity is represented as a function of scale, for four FDR levels: $0.01, 0.05, 0.1 ,0.2$. In addition to FDR control within scale, an omnibus test at $p<0.05$ was performed. Each column corresponds to a certain proportion of non-null hypothesis per family $\pi_1$ ($0\%,1\%,2\%,5\%,10\%$), and each row corresponds to a different combination of effect and sample size $N$ in $\{40,100\}$, $a^2$ in $\{0.1,0.2\}$, see text for details. The same clusters were used to simulate the time series and perform the tests.} 
}
\label{fig_sens_dep_multiscale_300_step_10_perc0}
\end{figure}

\begin{figure}[!ht]
\begin{center}
\includegraphics[width=\linewidth]{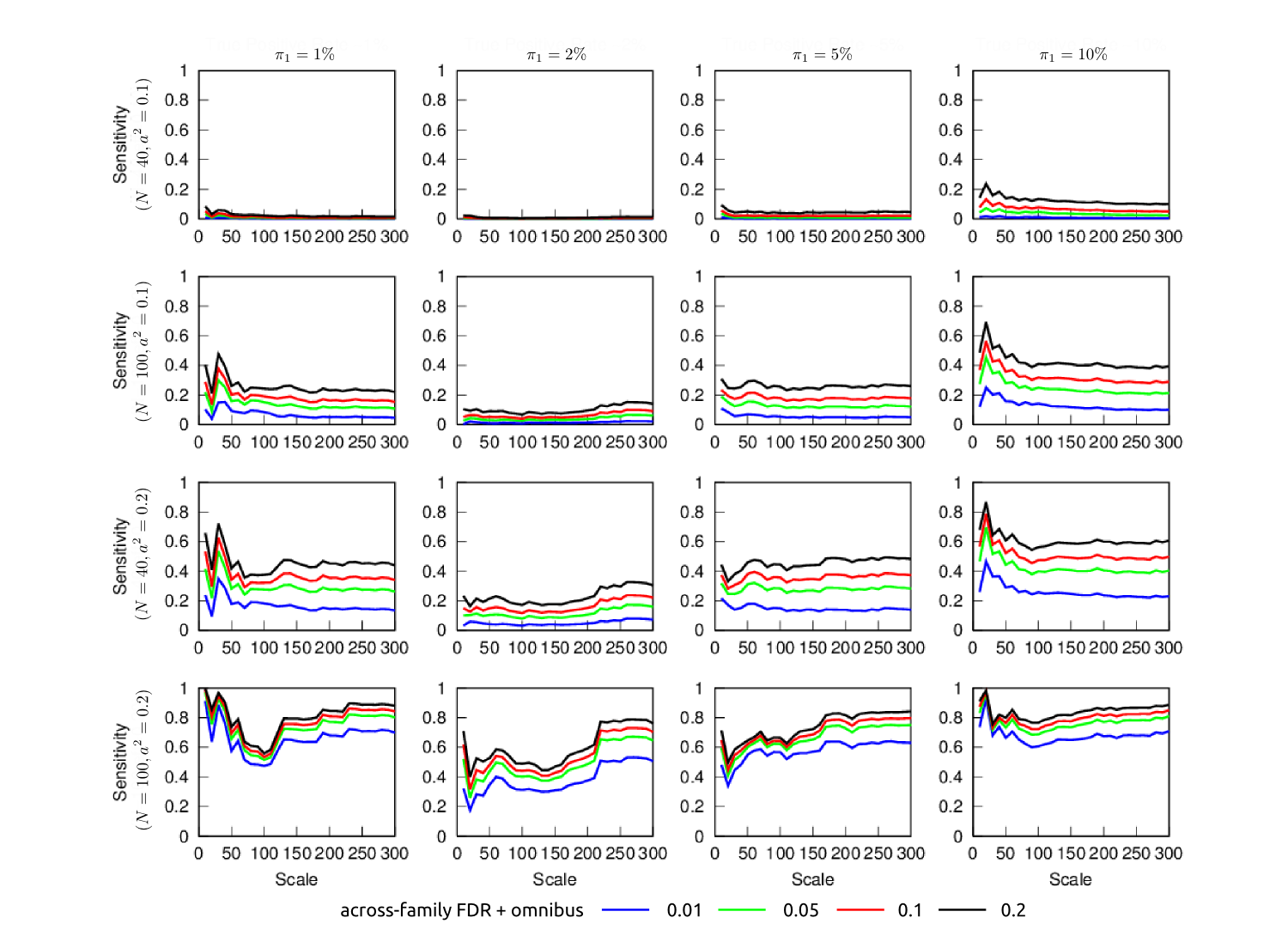}
\end{center}
\caption{
{\bf Sensitivity on simulations with dependent tests, when a $30\%$ mismatch is introduced between the true and test clusters ($K=30$, $L_k$ ranging from $55$ to $45150$, corresponding to the number of connections associated with a regular grid of scales covering 10 to 300 with a step of 10).} {The sensitivity is represented as a function of scale, for four FDR levels: $0.01, 0.05, 0.1 ,0.2$. In addition to FDR control within scale, an omnibus test at $p<0.05$ was performed. Each column corresponds to a certain proportion of non-null hypothesis per family $\pi_1$ ($0\%,1\%,2\%,5\%,10\%$), and each row corresponds to a different combination of effect and sample size $N$ in $\{40,100\}$, $a^2$ in $\{0.1,0.2\}$, see text for details. A 30\% mismatch was introduced between the true and the test clusters. The true positives used to estimate sensitivity were defined by the reference clusters for the simulation, regardless of the potential perturbation of these clusters prior to simulation to create a mismatch between true and test clusters. } 
}
\label{fig_sens_dep_multiscale_300_step_10_perc30}
\end{figure}

\paragraph{Sensitivity} When the true and test clusters perfectly matched, the sensitivity across scales followed a similar patterns in all scenarios: a decrease in sensitivity with increasing scales, although not as sharp as what was observed on simulations with independent tests, see Figure \ref{fig_sens_dep_multiscale_300_step_10_perc0}. This closely mirrored the large increase in effect size at low scales, due to averaging on clusters that perfectly matched the simulated signal, see Supplementary Figure \ref{fig_eff_size_dep_multiscale_300_step_10}. We noted that the simulation settings where departure from nominal levels were observed were also characterized by very low rate of discoveries, notably at high scale, with sensitivity below 0.1 for scales higher than 50 and falling to zero for scales higher than 150 (Figure \ref{fig_sens_dep_multiscale_300_step_10_perc0}, first row). By contrast, when introducing a $30\%$ mismatch between the true and test clusters, increases in sensitivity were observed across a wider range of low scales, e.g. $N=100$, $a^2=0.1$ and $\pi_1=10\%$, or even at high scales, e.g. $N=100$, $a^2=0.2$ and $\pi_1 = 2\%$ (Figure \ref{fig_sens_dep_multiscale_300_step_10_perc30}). This again reflected the more variable profiles of effect size as a function of scales across scenarios after the introduction of a mismatch between the true and test clusters. These simulations demonstrated the possibility to have increase in sensitivity as a function of scale, and that these gains would potentially be dependent on the effect size, the mismatch between the true/test clusters, as well as the sample size. These observations were made for a regular grid of scales, but were identical using the MSTEPS scales from the SCHIZO dataset (not shown).

\section{Application to real datasets}
\label{sec_real}

\subsection{Methods} 
\label{sec_real_methods}

\paragraph{Participants} The SCHIZO dataset was contributed by the Center for Biomedical Research Excellence (COBRE) to the 1000 functional connectome project\footnote{\url{http://fcon_1000.projects.nitrc.org/indi/retro/cobre.html}} \citep{Biswal2010}. The sample comprised 72 patients diagnosed with schizophrenia (58 males, age range = 18-65 yrs) and 74 healthy controls (51 males, age range = 18-65 yrs). The BLIND \citep{Collignon2011} and MOTOR \citep{Albouy2014} datasets were acquired at the Functional NeuroImaging Unit, at the Institut Universitaire de G\'eriatrie de Montr\'eal, Canada. Participants gave their written informed consent to take part in the studies, which were approved by the research ethics board of the Quebec Bio-Imaging Network (BLIND, MOTOR), as well as the ethics board of the Centre for Interdisciplinary Research in Rehabilitation of Greater Montreal (BLIND). The BLIND dataset was composed of 14 congenitally blind volunteers recruited through the Nazareth and Louis Braille Institute (10 
males, age range = 26-61 yrs) and 17 sighted controls (8 males, age range = 23-60 yrs). The MOTOR sample included 54 healthy young participants (33 males, age range = 19-33 yrs). 

\paragraph{Acquisition} Resting-state fMRI scans were acquired on a 3T Siemens TrioTim for all datasets. One single run was obtained per subject for either the SCHIZO or BLIND dataset while two runs were acquired in each subject for the MOTOR dataset, one immediately preceding and one following the practice on a motor task. For the SCHIZO dataset, 150 EPI blood-oxygenation level dependent (BOLD) volumes were obtained in 5 mns (TR = 2~s, TE = 29~ms, FA = 75\textdegree, 32 slices, voxel size = 3x3x4~mm$^3$, matrix size = 64x64), and a structural image was acquired using a multi-echo MPRAGE sequence (TR = 2.53~s, TE = 1.64/3.5/5.36/7.22/9.08~ms, FA = 7\textdegree, 176 slices, voxel size = 1x1x1 mm$^3$, matrix size = 256x256). For the BLIND dataset, 136 EPI BOLD volumes were acquired in 5 mins (TR = 2.2s, TE = 30~ms, FA = 90\textdegree, 35 slices, voxel size = 3x3x3.2 mm$^3$, gap = 25\%, matrix size = 
64x64), and a structural image was acquired using a MPRAGE sequence (TR = 2.3~s, TE = 2.91~ms, FA = 9\textdegree, 160 slices, voxel size = 1x1x1.2 mm$^3$, matrix size = 240x256). For the MOTOR dataset, 150 EPI volumes were recorded in 
6 mins 40 s (TR = 2.65s, TE = 30ms, FA = 90\textdegree, 43 slices, voxel size = 3.4x3.4x3 mm$^3$, gap = 10\%, matrix size = 64x64), and a structural image was acquired using a MPRAGE sequence (TR = 2.3~s, TE = 2.98~ms, FA = 9\textdegree, 176 slices, voxel size = 1x1x1 mm$^3$, matrix size = 256x256).

\paragraph{Motor task} Between the two rest runs of the MOTOR experiment, subjects were scanned while performing a motor sequence learning task with their left non-dominant hand. 14 blocks of motor practice were interspersed with 15s rest epochs. Motor blocks required subjects to perform 60 finger movements, ideally corresponding to 12 correct five-element finger sequence. The duration of the practice blocks decreased as learning progressed. It should be noted that the effect of motor learning per se on the subsequent rest run could not be distinguished from that of a mere motor practice/fatigue effect in the present experimental setting.

\paragraph{Preprocessing} Each fMRI dataset was corrected for inter-slice difference in acquisition time and the parameters of a rigid-body motion were estimated for each time frame. Rigid-body motion was estimated within as well as between runs, using the median volume of the first run as a target. The median volume of one selected fMRI run for each subject was coregistered with a T1 individual scan using Minctracc \citep{Collins1997}, which was itself non-linearly transformed to the Montreal Neurological Institute (MNI) template \citep{Fonov2011a} using the CIVET pipeline \citep{Ad-Dabbagh2006}. The MNI symmetric template was generated from the ICBM152 sample of 152 young adults, after 40 iterations of non-linear coregistration. The rigid-body transform, fMRI-to-T1 transform and T1-to-stereotaxic transform were all combined, and the functional volumes were resampled in the MNI space at a 3 mm isotropic resolution. The “scrubbing” method of \cite{Power2012}, was used to remove the volumes with excessive 
motion (frame displacement greater than 0.5 mm). A minimum number of 60 unscrubbed volumes per run, corresponding to $\sim 180$ s of acquisition, was then required for further analysis. For this reason, some subjects were rejected from the subsequent analyses: 16 controls and 29 schizophrenia patients in the SCHIZO dataset (none in either the BLIND or MOTOR datasets). The following nuisance parameters were regressed out from the time series at each voxel: slow time drifts (basis of discrete cosines with a 0.01 Hz high-pass cut-off), average signals in conservative masks of the white matter and the lateral ventricles as well as the first principal components (95\% energy) of the six rigid-body motion parameters and their squares \citep{Giove2009}. The number of confounds regressed from the individual time series ranged from 12 to 18 for the MOTOR sample, from 11 to 15 for the BLIND sample, and from 10 to 17 for the SCHIZO sample. The fMRI volumes were finally spatially smoothed with a 6 mm isotropic Gaussian blurring kernel. Note that the preprocessed fMRI time series for the COBRE experiment have been made publicly available\footnote{\url{http://figshare.com/articles/COBRE_preprocessed_with_NIAK_0_12_4/1160600}}. 

\paragraph{Multiscale parcellation} Brain parcellations were derived using BASC separately for each dataset, while pooling the patient and control groups in the SCHIZO and BLIND datasets, and runs in the MOTOR dataset. The BASC used 100 replications of the clustering of each individual time series, using circular block bootstrap, and 500 replications of the group clustering, using regular bootstrap. The functional group clusters were first generated on a fixed regular grid, from $10$ to $300$ clusters with a step of $10$, identical for all three real datasets, and with identical numbers of individual, group, and final (consensus) clusters. The MSTEPS procedure was then implemented to select a data-driven subset of scales approximating the group stability matrices up to 5\% residual energy, through linear interpolation over selected scales. The number of individual clusters were selected in the fixed grid above, yet the number of group and final clusters were searched in an interval of $\pm 20\%$ of the individual scales, such that the final scales were not constrained in the range $10$ to $300$. Note that the number of scales itself was selected by the MSTEPS procedure in a data-driven fashion, and that the number of individual, group and final (consensus) number of clusters were not necessarily identical. 

\paragraph{General linear model} For all GLM analyses, the covariates included an intercept, the age and sex of participants as well the average frame displacement of the runs involved in the analysis (two covariates of frame displacement were used in the MOTOR dataset, one per run). The contrast of interest was on a dummy covariate coding for the difference in average connectivity between the two groups for SCHIZO and BLIND, and on the intercept (average of the difference in connectivity pre- and post-training) for the MOTOR dataset. Note that for the motor dataset the difference in connectivity between the second run and the first run was entered in the group-level GLM, in place of the individual connectome. All covariates except the intercept were corrected to a zero mean.

\paragraph{Modeling assumptions} The parametric GLM relies on a series of assumptions, most critically the normality of distribution of the residuals of the tests, and the homoscedasticity of residuals, i.e. equal variance across subjects. For each connection and each contrast, the normality of distribution for the residuals of the regression was tested with a composite test\footnote{as implemented in the \texttt{swtest.m} procedure \url{http://www.mathworks.com/matlabcentral/fileexchange/13964-shapiro-wilk-and-shapiro-francia-normality-tests/content/swtest.m}, retrieved on 12/2014.}: Shapiro-Francia for platykurtic distributions and Shapiro-Wilk for leptokurtic distributions \citep{Royston1993}. A test for homoscedastic residuals was also implemented using the procedure of \cite{White1980}, where all variables as well as their two-way interactions (including squared variables) were regressed against the square of the residuals, and a $F$ test was performed on the combination of all exploratory variables. A $p$ value was generated at each connection, both for the normality and the homoscedasticity tests, for the highest resolution selected by MSTEPS, and multiple comparisons across all connections were corrected with the FDR-BH procedure ($q<0.05$). In addition to the MOTOR, BLIND and SCHIZO datasets, the Cambridge dataset previously used in the simulations was also employed here. The GLM only included an intercept and an arbitrary group difference, for different sample sizes ($N \in \{40, 100, 180\}$), in order to investigate how the testing of assumptions behaved for different sample sizes. 

\subsection{Results} 
\label{sec_res_real}
\paragraph{Modeling assumptions} No tests of departure from normality passed correction for multiple comparison using FDR-BH at $q<0.05$. However, some trends towards significance were observed in all datasets, in particular with a large sample size. For a threshold of $p<0.05$, uncorrected for multiple comparisons, the normality hypothesis was rejected for $9\%$, $6.8\%$ and $11\%$ of connexions, for the MOTOR, BLIND and SCHIZO experiments, respectively (Supplementary Figure \ref{fig_test_gaussian}). 
\par
No test for heteroscedasticity survived a correction for multiple comparisons with FDR-BH at $q<0.05$, and there was no apparent trend. At $p<0.05$, the homoscedasticity hypothesis was rejected for $3.4\%$, $4.2\%$ and $7.4\%$ in the MOTOR, BLIND and SCHIZO experiments, respectively. The trends observed for heteroscedasticity testing were similar to those observed in the Cambridge dataset, using random subgroups that are thus in fact homoscedastic (Supplementary Figure \ref{fig_test_hetero}). 

\begin{figure}[!ht]
\begin{center}
\includegraphics[width=\linewidth]{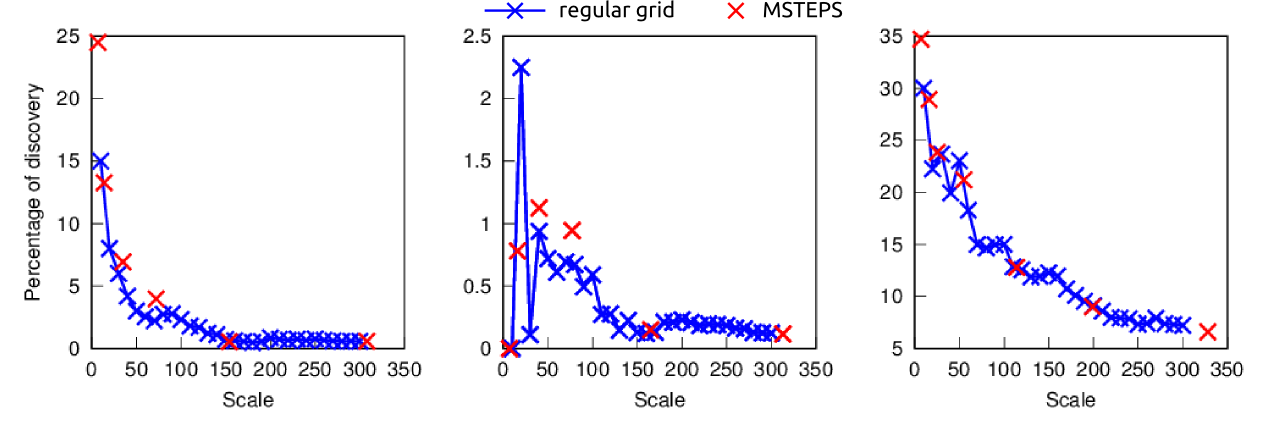}
\end{center}
\caption{
{\bf Percentages of discovery as a function of scales.} {Plots show the percentage of discovery for the MOTOR, BLIND and SCHIZO contrasts. The blue curve represent the scales selected on a regular grid, from 10 to 300 with a step of 10, and the red crosses show the scales selected by the MSTEPS procedure (see text for details).} 
}
\label{fig_perc_disc}
\end{figure}

\paragraph{Multiscale discoveries} The MSTEPS procedure selected 6 scales for the MOTOR and BLIND samples, and 7 on the SCHIZO sample, ranging from 7 to 300+, see Table \ref{tab_msteps} for multi-level scale parameters. The MSPC detection generated maximal percentages of discoveries at low scales for the three datasets (Figure \ref{fig_perc_disc}). Using a grid from 10 to 300 scales with a step of 10, peak discoveries were detected at scale 10 for the SCHIZO and MOTOR contrasts, and scale 20 for the BLIND contrast. Peak percentages of discoveries were 30\%, 2.3\% and 15\%, for the SCHIZO, BLIND and MOTOR contrasts, respectively. The omnibus test was significant ($p<0.05$) for all three contrasts, whether using a large grid of 30 scales or the 6-7 scales identified with MSTEPS. The overall trend was that the rate of discoveries decreased as the number of parcels increased, with the largest discovery rate found below scale 50, followed by a notable plateau from 50 to 100 clusters. These relationships between discovery rate and scales shared similarities with the sensitivity plots observed on simulations (Figures \ref{fig_multiscale_msteps_schizo}, \ref{fig_sens_dep_multiscale_300_step_10_perc30}). While the absolute percentages of discoveries were quite different for the three datasets, the relative changes in discovery rate as a function of scale were thus rather similar. 

\begin{figure}[!ht]
\begin{center}
\includegraphics[width=0.75\linewidth]{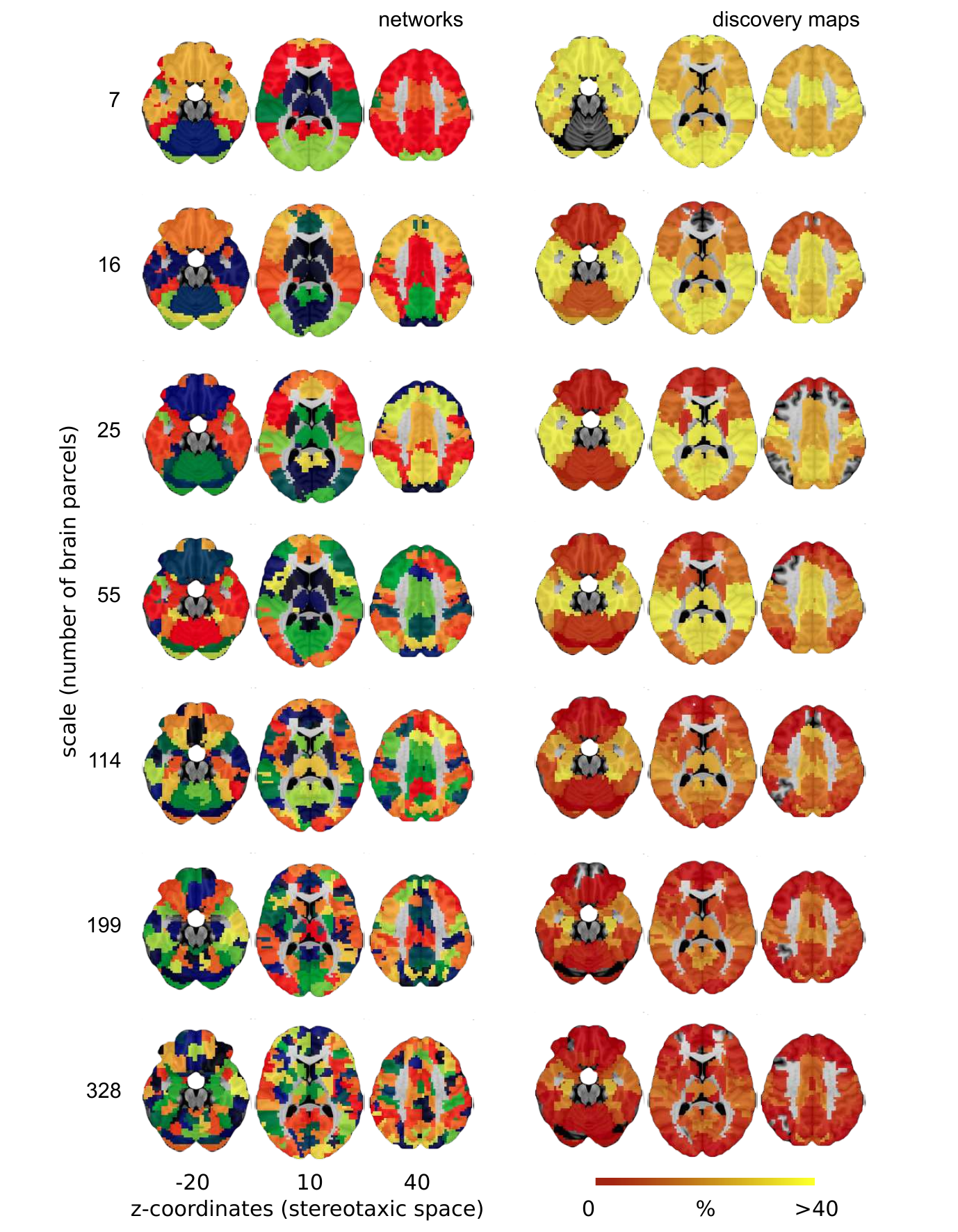}
\end{center}
\caption{
{\bf MSTEPS parcels and percentage of discovery maps in the SCHIZO contrast, in volumetric space.} { Networks show the functional brain parcellations for the 7 MSTEPS scales. Corresponding percentage discovery maps show the percentage of connections with a significant effect, for each brain parcel. MNI coordinates are given for representative slices superimposed onto the MNI 152 non-linear template.} 
}
\label{fig_perc_disc_sz}
\end{figure}

\begin{figure}[!ht]
\begin{center}
\includegraphics[width=\linewidth]{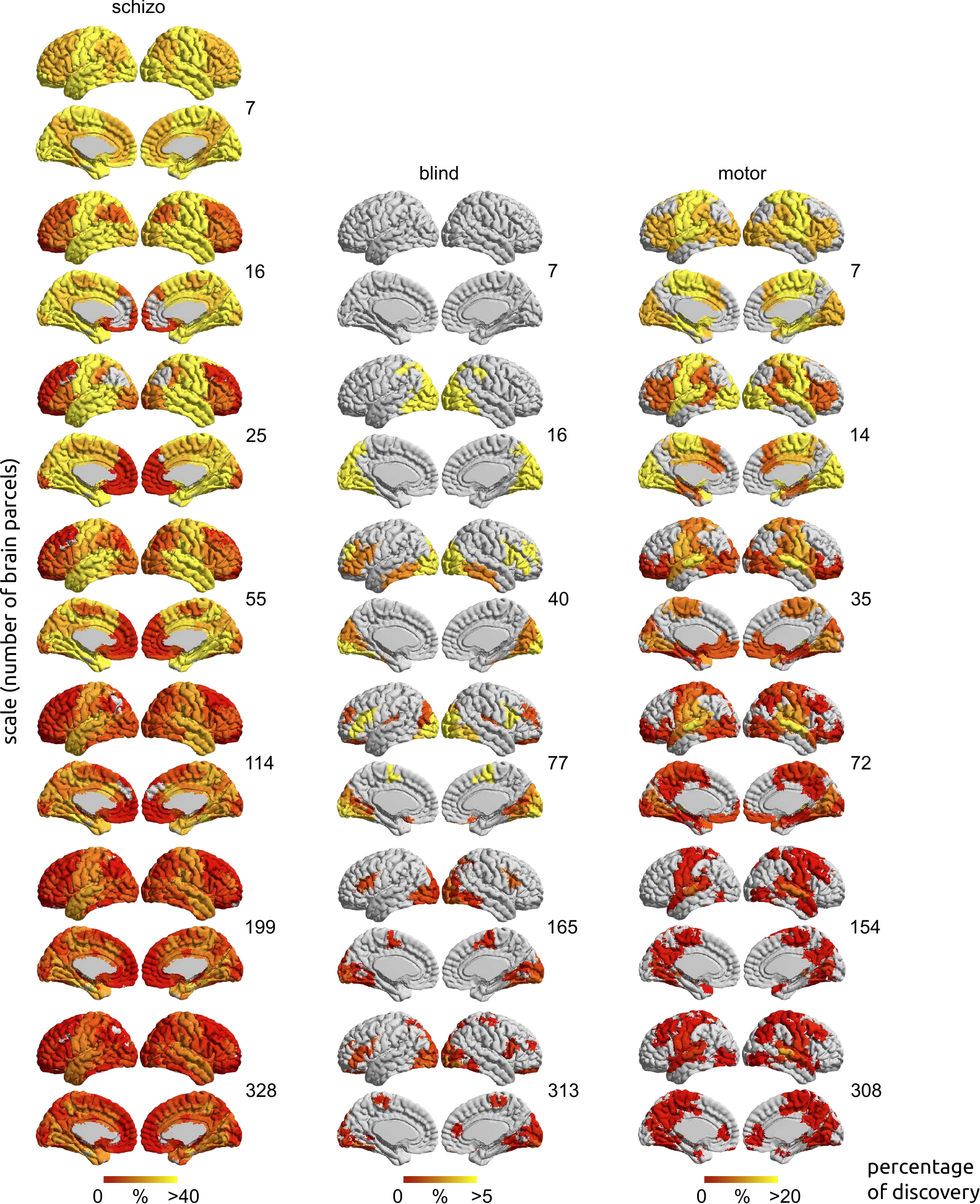}
\end{center}
\caption{
{\bf Percentage of discovery maps in the three real datasets for all MSTEPS scales.} {Surface maps show the percentage of connections with a significant effect, for each brain parcel, in respectively the SCHIZO, BLIND and MOTOR contrasts. Maps are projected onto the MNI 2009 surface. See Figure \ref{fig_perc_disc_sz} for volumetric representations showing results at the subcortical level in the SCHIZO contrast.}
}
\label{fig_perc_disc_maps}
\end{figure}

\paragraph{Spatial distribution of significant discoveries} Discovery percentage maps revealed which parcels were associated with the largest proportion of significant connections for any given parcel, see Figure \ref{fig_perc_disc_sz} for a representation of the BASC multiscale parcels and associated discovery percentage maps for the SCHIZO analysis. For each contrast, results were shown for all 6-7 scales extracted with the MSTEPS procedure. The areas showing maximal percentage of discoveries were quite different for the three datasets (Figure \ref{fig_perc_disc_maps}). Widespread effects were observed for the SCHIZO dataset at both cortical and subcortical levels (see also Figure \ref{fig_perc_disc_sz}, for a volumetric representation). Parcels with the highest discovery rate were found in the temporal cortex, the medial temporal lobe, the anterior cingulate cortex and the basal ganglia. The BLIND contrast revealed more localized effects, in the occipital cortex and to a lesser extent in the temporal and frontal cortices. Finally, the MOTOR contrast identified significant effects within an extended visuomotor cortico-subcortical parcel. 

Despite the overall higher rate of discoveries for low scales, below 50, the spatial distributions of discoveries were fairly consistent across scales. It was also interesting to note that low scales did not always provide the highest discovery rate for a given brain parcel. For instance, the proportion of connections showing a significant effect in the basal ganglia for the SCHIZO contrast became maximal at scale 55, once the thalami were isolated as a single parcel (Figure \ref{fig_perc_disc_sz}). As another example, the dorsolateral prefrontal cortex only showed a significant effect in the BLIND contrast for functional brain parcellations above scale 40 (Figure \ref{fig_perc_disc_maps}).

\begin{figure}[!ht]
\begin{center}
\includegraphics[width=\linewidth]{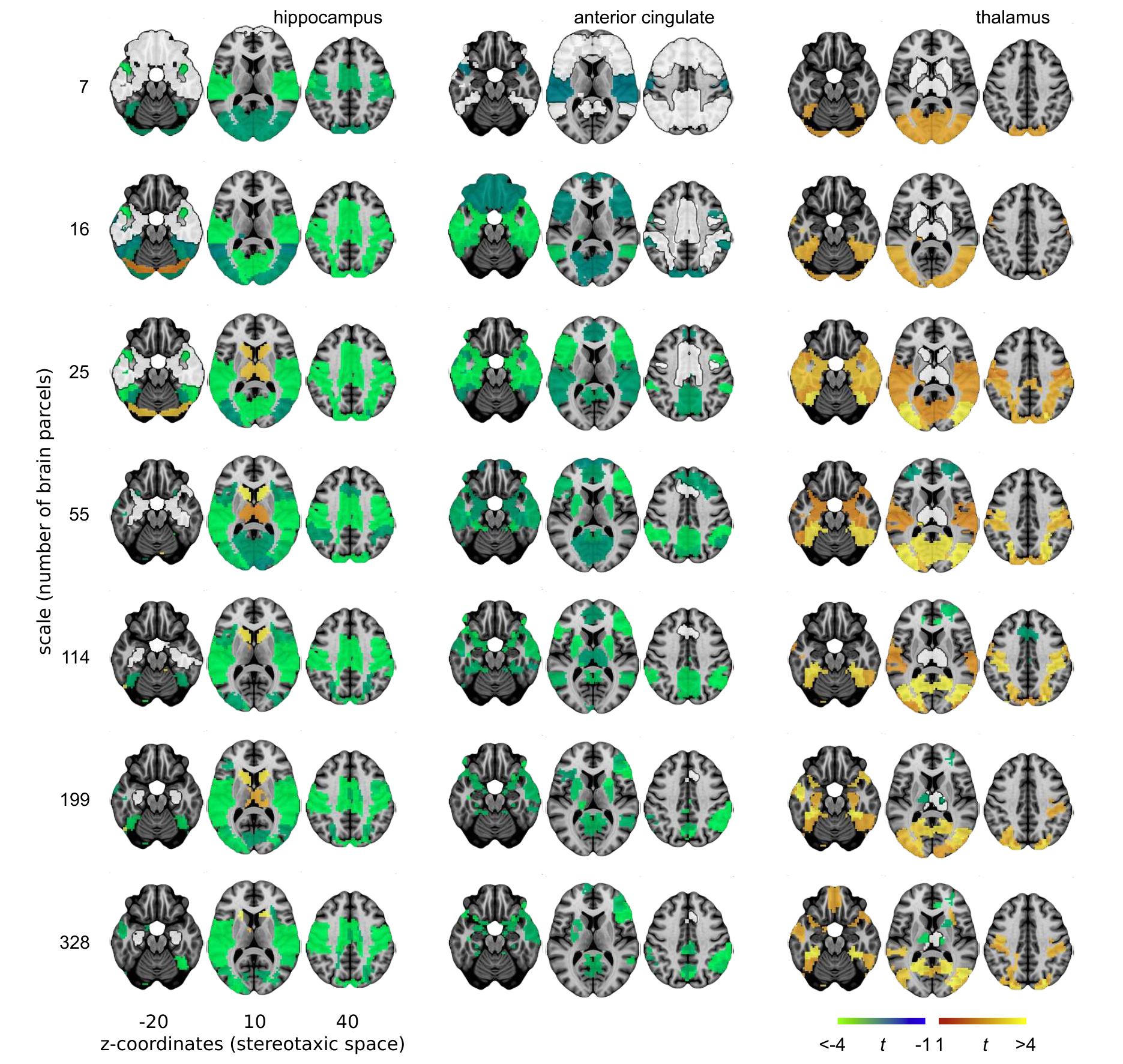}
\end{center}
\caption{
{\bf  Group FDR-corrected $t$-test maps in the SCHIZO dataset, in volumetric space.} {$T$-test maps showed significant alterations ($q<0.05$ for FDR-BH) in functional connectivity (decreases and increases) in schizophrenia for the 7 MSTEPS scales and several seeds. The seed that included the hippocampus, the anterior cingulate and the thalami were shown as stroke white parcels at all scales. Intra-parcel changes in connectivity were thus not shown for seeds (e.g., decreased connectivity within the basal ganglia). The $z$ MNI coordinates were given for representative slices superimposed onto the MNI 152 non-linear template.} 
}
\label{fig_eff_maps}
\end{figure}

\paragraph{Seed-based maps of $t$-statistics} The maps of discovery rate did not characterize which specific connections were identified as significant for each parcel, nor the direction of the effect (i.e. an increase vs a decrease in connectivity). We illustrated how these questions can be explored using the SCHIZO dataset, as it showed widespread changes in functional connectivity. The percentage of discovery maps were used to select a number of seed parcels of interest, i.e. showing maximal effects (Figure \ref{fig_eff_maps}). Parcels selected at the highest scales corresponded to the hippocampus, anterior cingulate cortex and thalamus. Corresponding parcels for lower scales were selected based on their maximal overlap with the parcels chosen at the highest scales. For instance, the most distributed parcel encompassing the hippocampus at scale 7 covered the whole medial temporal lobe, the temporal pole and ventral prefrontal cortex. For each brain parcel, a FDR-corrected $t$-test map associated with the contrast of interest was generated. These $t$-test maps revealed that the alterations in functional coupling in schizophrenia essentially took the form of a decrease in connectivity for the hippocampus and associated regions as well as for the anterior cingulate cortex and its associated parcel. By contrast, the thalamus and basal ganglia showed an increase in functional connectivity with the occipital cortex, beyond decreased connectivity within the basal ganglia. 

\begin{figure}[!ht]
\begin{center}
\includegraphics[width=0.5\linewidth]{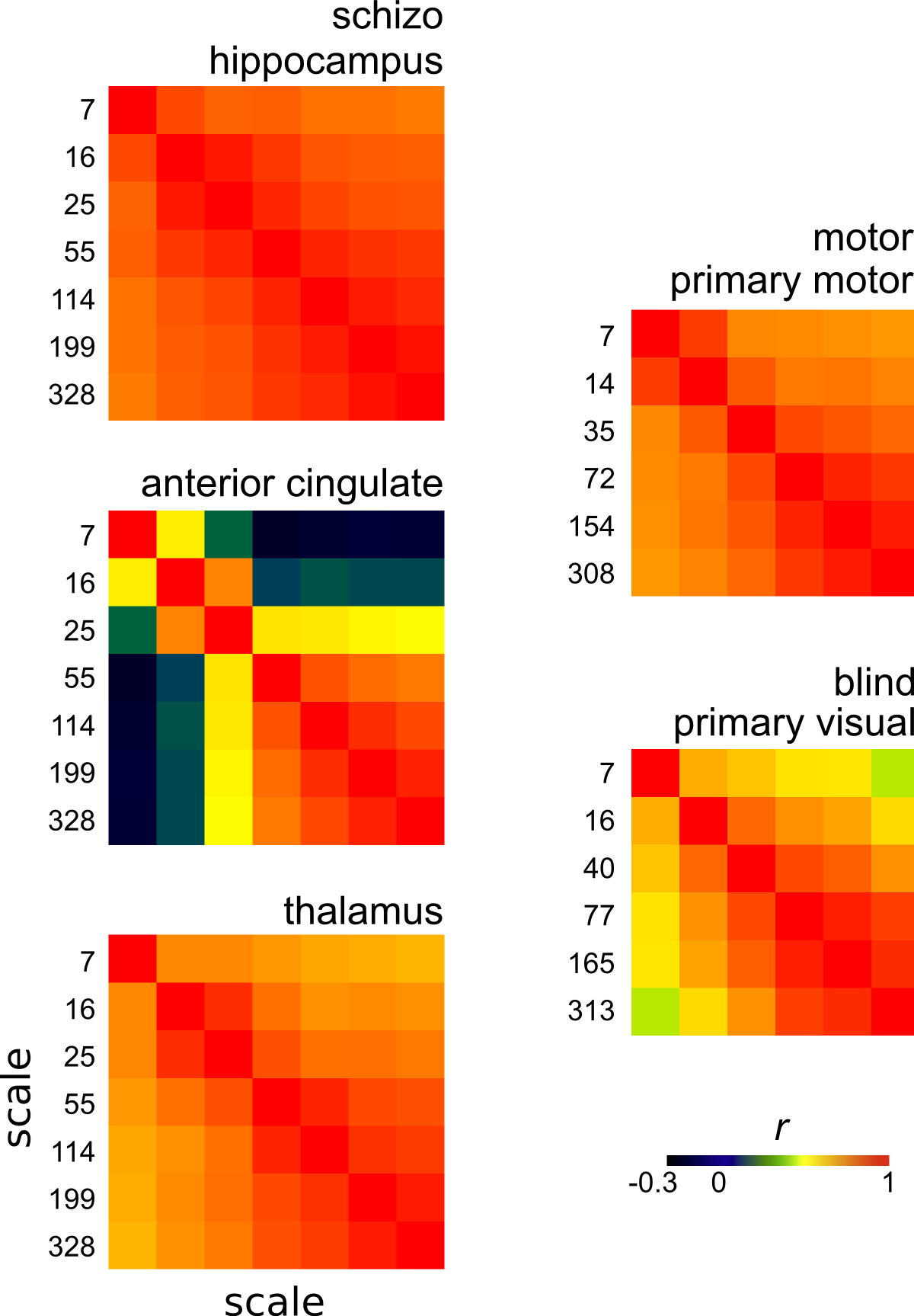}
\end{center}
\caption{
{\bf Correspondence of effects maps across scales for the three real datasets.} {Correlation matrices show pairwise comparisons between 7 or 6 MSTEPS scales of the effects maps for three selected seeds in the SCHIZO dataset and one a priori seed in each of the BLIND and MOTOR dataset.}
}
\label{fig_eff_multiscale}
\end{figure}

\paragraph{Impact of scale on statistical maps} While visual exploration of the $t$-test maps in the SCHIZO dataset revealed similarities of the effects across scales, it also highlighted some specificities. High scales indeed proved in some cases to be additionnally informative compared to low scales, despite decreased overall detection rate. For instance, the parcel centered on the hippocampus was seen to be more positively connected with the thalamus and caudate nucleus in schizophrenia only when the ventral prefrontal cortex was not part of the parcel (Figure \ref{fig_eff_maps}). As another exemple, the thalamus showed increased connectivity with a large sensorimotor cortical parcel at scale 25 and above only, when it was not part of the same parcel as the putamen. Furthemore, the thalamus only showed a significant decrease in connectivity with the dorsolateral prefrontal cortex at scale 55 and above, when isolated as a single parcel rather than smoothed out inside the basal ganglia.
\par
We more formally tested the level of correspondence of the effects across scales for the three seeds listed above in the SCHIZO dataset, as well as for seeds matching our a priori in the BLIND and MOTOR datasets, respectively located in the right primary visual cortex and the left primary motor cortex. Pairwise comparisons between spatial effect maps across scales mostly revealed positive correlation values in all three datasets and for all seeds (Figure \ref{fig_eff_multiscale}). Correlations for the three seeds investigated in the SCHIZO contrast were as follows: hippocampus (mean, standard deviation, minimum, maximum = 0.86, 0.06, 0.76, 0.97), anterior cingulate (0.41, 0.39, -0.20, 0.93), and thalamus (0.78, 0.09, 0.64, 0.94). High correlations were always observed when comparing high scales (above scale 55) between them. Comparisons between low and high scales remained associated with high correlations values for two out of the three seeds, namely the hippocampus and thalamus. However, results for the anterior cingulate demonstrated that a low correspondence between low and high scales was possible. Results for the seeds in the BLIND (0.71, 0.15, 0.46, 0.93) and MOTOR (0.80, 0.08, 0.70, 0.95) datasets further supported the general conclusions drawn from the SCHIZO dataset.

\section{Discussion}
\label{sec_discussion}

\subsection{False-discovery rate within and across scales} This work investigated empirically how the scale impacted the results of a GLM analysis on connectomes. We first assessed the validity of a GLM analysis at a single scale. On the three real datasets, there was no sign of substantial departure from the assumptions of a basic parametric GLM. Importantly, the control of the effective FDR within scale on simulations was exact or conservative, including the Cambridge ``global null'' experiment where real random subgroups were compared. Our empirical evaluation thus supports the validity of parametric GLM analysis of connectomes at a single scale. 
\par
We also investigated the specificity of MSPC analysis across scales. In most simulation experiments, the control of the FDR within each scale also implied a tight control of the FDR across scales. Under the global null hypothesis, the omnibus test precisely controlled the FDR (or, equivalently, the FWE). The omnibus test had little to no impact in scenarios with many true non-null hypothesis, where the FDR across scales was conservatively controlled. The only simulations where a deviation of the FDR accross scales from nominal levels was observed (up to 0.09 for a nominal level at 0.05) typically resulted in very few findings, especially at high scales (above 50). Considering that the FDR was always appropriately controlled within scale, our overall conclusion was that, for scales matching those used in our simulation experiments, the inflation of FDR across scales was not a concern as long as the omnibus test was rejected at an appopriate FWE level. 

\subsection{Sensitivity across scales}
We found some convergent evidence regarding the sensitivity of MSPC in our simulation and real data experiments. First, on simulations with independent tests, the sensitivity decreased sharply with scale, to reach a plateau around scale 50. This behaviour appeared to be a consequence of multiple testing in the FDR-BH procedure, as the proportion of true non-null hypothesis and the effect size were strictly maintained at a constant level across scales. In our simulations of dependent tests, the scale directly impacted the effect size, as some test clusters matched better the underlying simulated signals than others. In this setting, marked increase (up to 0.3) in sensitivity were observed at certain scales, sometimes quite high (200+). In summary, scale-dependent gains in sensitivity were observed and appeared to be simultaneously driven by the match between the test clusters and the underlying signal as well as a mechanic effect of multiple comparisons in the FDR-BH procedure. 
\par
On real data, we investigated the discovery rate, i.e. the proportion of significant connections in a connectome. The same behaviour was observed on the three datasets: highest discovery rate at low scale (below 50 parcels), a plateau from 50 to 100 and a low asymptote above 100-200 parcels. This profile resembled most closely the sensitivity results from simulations with independent tests, and may reflect some intrinsic property of the FDR-BH procedure. In particular, scales larger than 100, routinely used with the AAL template \citep{Tzourio-Mazoyer2002}, was systematically associated with a much smaller discovery rate than lower scales (below 50). 
\par
In our three experiments, we did not identify strong discrepancies between statistical maps generated at different scales, consistent with the observations of \cite{Shehzad2014}. This supports the idea that, in many cases a GLM analysis at a single scale may be enough to summarize the results of a GLM on connectomes. We observed that effects in some brain parcels were better captured at particular scales. For example, the difference in thalamic connectivity in the SCHIZO analysis was better seen at scale 55 and above, where the thalami were clustered in one parcel rather than agregated with the putamen and caudate nuclei.
 
\subsection{Selection of an ``optimal'' scale}
The control of the FDR across scales, which is central to the proposed MSPC approach, is only relevant if the results of a GLM-connectome analysis are investigated and reported at mutiple scales. It would not be advisable to run a MSPC and then simply report results at the scale with the highest discovery rate. There would be no guarantee that the FDR would be controlled for a scale that was selected precisely because of a high associated discovery rate, a classic case of circular analysis. If a single scale were considered in an analysis, the selection of this scale should be either set a priori or selected through MSPC on an independent dataset. However, as we noted previously, our findings on three experiments follow an overall trend that suggests that a single scale below 50 would likely be a better default choice than a template at scale 100+, such as AAL \citep{Tzourio-Mazoyer2002}, in terms of discovery rate. 

\subsection{Findings on real datasets}
The effects found on the real datasets were consistent with the existing literature. First, schizophrenia has been defined as a dysconnectivity syndrome, with aberrant functional interactions between brain regions being a core feature of this mental illness \citep[for reviews, see ][]{Calhoun2009,Pettersson-Yeo2011,Fornito2012}. As shown here for two out of three parcels, and observed for other unreported brain parcels, widespread decrease in connectivity was observed in patients, with the addition of more localized increases in connectivity. The prominence of decreases in connectivity in the temporal lobe, hippocampus and anterior cingulate cortex, amongst other regions, is well supported by previous studies \citep{Williamson2012}. Similarly, increased connectivity between the thalamus and sensorimotor cortex but decreased connectivity with striatal and prefrontal regions has been reported before \citep{Anticevic2014}. Second, resting-state fMRI studies have previously shown that congenital blindness is associated with a reorganization of the interactions between the occiptial cortex and other parts of the brain, in particular the auditory and premotor cortices \citep{Liu2007,Qin2013,Qin2014}, consistent with our findings. Finally, our results are in agreement with the observation that brain activity at rest is modulated by previous intensive motor practice \citep{Albert2009,Vahdat2011,Sami2014}. Even in the absence of a definite ground truth on these real life applications, our findings thus had good face validity, and suggested that MSPC could be successfully applied to a variety of clinical or experimental conditions. 

\subsection{Comparison with other methods}
In this work, we only controlled for multiple comparisons within each scale using the FDR-BH method, yet several alternative methods have recently been proposed. \cite{Shehzad2014} developped a multivariate test that applies on a region-to-brain connectivity map, called multivariate distance matrix regression (MDMR). This procedure would be used to screen for promising seed-based connectivity maps worthy to explore in a subsequent, independent analysis. The MDMR approach effectively performs one test per parcel, instead of one test per connection, and thus greatly alleviates the multiple comparison problem. It does not however provide a control of statistical risk at the level of single connections. \cite{Zalesky2010} proposed to use uncorrected threshold on the individual $p$-values, but then to identify to which extent the connections that survive the test are interconnected. This extent measure is compared against what could be observed under a null hypothesis of no association, implemented through permutation testing. This approch, called Network-Based Statistics (NBS), is the connectome equivalent to the ``cluster-level statistics'' used in SPMs. The NBS only offers a loose control of false-positive rate at the level of a single connection, but can be used to reject the possibility that a group of significant findings could be observed by chance in the FWE sense. Both the MDMR and NBS do not offer control of the FDR within scale, and thus cannot be used with the MSPC approach based on the argument that the FDR control extends across scales. Some of these methods have been compared in a variety of scenarios, e.g. FDR-BH and NBS \citep{Zalesky2012}, but at a fixed scale. Based on the observation of important variations in sensitivity across scales for the FDR-BH, an important avenue of future work would be to investigate how MDMR, NBS and FDR-BH compare at different scales. Of note is the recent work of \cite{Meskaldji2014}, which combines two scales to peform connectome-wide testing: the low scale is used to screen for promising groups of intra- or inter-parcel connections, and the tests at high scale are re-weighted based on that screening. The weights can be adjusted to ensure control of the FDR across the connectome. This alternative approach to multiscale testing is limited to two scales, but may provide additional statistical power compared to MSPC as the two scales are analyzed in coombination rather than separately. 

\subsection{Beyond scale selection: choice of the brain parcellation}
Although the impact of the number of parcels on CWAS sensitivity has been extensively investigated in this work, we only briefly examined how the choice of parcels, and not just their number, could impact sensitivity. We could, for example, have used random parcellations, like \citep{Zalesky2010a}, a parcellation based on anatomical landmarks such as the AAL atlas \citep{Tzourio-Mazoyer2002}, or a functional parcellation with spatial connexity constraints \citep{Craddock2012}. From our results on simulations, it seems clear that dramatic differences in statistical power can be achieved at a given spatial scale, if a set of parcels is best adapted to the spatial distribution of an effect. The work of \cite{Craddock2012} suggested that functional brain parcels are more homogeneous than anatomical parcels. We believe that important improvement in sensitivity could be gained from the optimization of the parcellation scheme, rather than scale, and this represents an important avenue for future research. Following an idea initially explored in \citep{Thirion2006a}, it may even be possible to relax the constraint of identical parcels across subjects, by matching different individual-specific parcels and use this correspondence to run group-level CWAS analysis.

\section{Conclusion}
Our overall conclusion is that the MSPC method is statistically valid (specific) and has the potential to identify biologically plausible associations in a variety of experimental conditions. An analysis at a single scale with less than 50 parcels appears as a reasonable default option, likely to have a sensitivity superior to the common approach using 100+ brain parcels in many settings. Multiscale analysis still have the potential to identify specific effects in small parcels. The MSPC method is available in the NIAK package\footnote{\url{nitrc.org/projects/niak}} \citep{Bellec2011}, a free and open-source software that runs in matlab and GNU octave, and we also released a set of multiscale functional brain parcellations  \footnote{\url{http://figshare.com/articles/Group_multiscale_functional_template_generated_with_BASC_on_the_Cambridge_sample/1285615}}. 

\section{Acknowledgments}
Parts of this work were presented at the 2012 and 2013 annual meetings of the organization for human brain mapping, as well as the International Conference on Resting-State Connectivity 2012 (Magdeburg). The authors are grateful to the members of the 1000 functional connectome consortium for publicly releasing the ``Cambridge'' and ``COBRE'' data samples, as well as Dr Shehzad for valuable feedback. The computational resources used to perform the data analysis were provided by Compute Canada\footnote{\url{https://computecanada.org/}} and CLUMEQ\footnote{\url{http://www.clumeq.mcgill.ca/}}, which is funded in part by NSERC (MRS), FQRNT, and McGill University. This project was funded by NSERC grant number RN000028, a salary award from ``Fonds de recherche du Qu\'ebec -- Sant\'e'' to PB as well as a salary award by the Canadian Institute of Health Research to PO.

\section*{References}

\bibliographystyle{elsarticle-harv}

\clearpage
\appendix

\section{Functional connectomes}
\label{app_connectome}
Let $\{ \mathcal{P}_i, i=1\dots R\}$ be a partition of the brain, i.e. $R$ parcels such that any voxel in the grey matter belongs to one and only one a parcel. The number of parcels $R$ is the (spatial) scale of the partition. For an fMRI dataset with $T$ time samples, the average time series $\mathbf{w}_{i}$ (vector of length $T$) is generated for each parcel $\mathcal{P}_î$. These average time series are then used to generate a $R\times R$ matrix of functional connectivity $\mathbf{Y}=(y_{i,j})_{i,j=1}^R$:
\begin{equation}
 \label{eq_R}
 y_{ij} = F\left(\textrm{corr}(\mathbf{w}_i,\mathbf{w}_j)\right), \quad \textrm{with}\,F(r) = \frac{1}{2}\log\left(\frac{1-r}{1+r}\right),
\end{equation}
where $\textrm{corr}$ is Pearson's linear correlation coefficient and $F$ is the Fisher's transform. The Fisher's transform is used to stabilize the variance of the estimated correlation coefficient \citep{Anderson1958}. This measure of between-parcel connectivity was used for $i\neq j$, but we also included a measure of within-cluster average functional connectivity, which uses the voxel-level time series $\mathbf{w}_v$:
\begin{equation}
y_{ii} = F\left(\frac{1}{\#\mathcal{P}_i(\#\mathcal{P}_i-1)}\sum_{v,v'\in\mathcal{P}_i,v\neq v'} \textrm{corr}(\mathbf{w}_v,\mathbf{w}_{v'})\right).
\end{equation}

\section{Ordinary least square GLM estimation}
\label{app_glm}
The independent and homoscedastic assumption means that the coefficients of $\mathbf{E}$ are independent from each other and that for each connection $l$, the $(e_{n,l})_{n=1}^N$ coefficients are identically distributed with a zero mean and variance $\sigma_l^2$. In this context, the maximum likelihood (ordinary least-squares) estimator of $\mathbf{B}$ is:
\begin{equation}
 \label{eq_lse}
 \hat{\mathbf{B}} = (\mathbf{X}'\mathbf{X})^{-1}\mathbf{X}'\mathbf{Y},
\end{equation}
and the estimation of the variance of the noise is:
\begin{equation}
 \label{var_noise}
 \hat{\sigma}_l^2 = \frac{1}{N-C}\sum_{n=1,\dots,N} \hat{e}_{n,l}^2, \quad \textrm{with} \quad \hat{\mathbf{E}} = \mathbf{Y}-\mathbf{X}\hat{\mathbf{B}}.
\end{equation}
For each covariate $c$, the vector $(\hat{\beta}_{c,l})_{l=1}^L$ is a vectorized connectome of statistical parameters, quantifying the modulation of each connection $l$ by the covariate $c$. This is a direct generalization of the concept of a SPM that has been widely used in task-based fMRI analysis. Each column of the statistical parametric connectome is a actually a SPM at the parcel level (instead of the more standard voxel level), testing the modulation of the functional connectivity of a given seed region with the rest of the brain by the covariate of interest. It is possible to test the signficance of each element of the statistical parametric connectome $\hat{\beta}_{c,l}$ against the null hypothesis $(\mathcal{H}_0)$ of no association (i.e. $\beta_{c,l} = 0$), using a $t$-test:
\begin{equation}
 t_l = (\delta_c'\hat{\mathbf{B}}_l)\left(\hat{\sigma}_l\sqrt{\delta_c'(\mathbf{X}'\mathbf{X})^{-1}\delta_c}\right)^{-1},
\end{equation}
where $\delta_c$ is a contrast (column) vector, with $\delta_{d}$ equals 1 for $(d=c)$ and 0 otherwise. Under $(\mathcal{H}_0)$, the quantity $t_l$ follows a Student's $t$ distribution with $N-C$ degrees of freedom. By comparing $t_l$ with the cumulative distribution function $g_{N-C}$ of the Student's distribution, it is possible to derive the bilateral probability of observing $t_l$ under $(\mathcal{H}_0)$:
\begin{equation}
 p_l = 2 \left( 1-g_{N-C}(|t_l|) \right) .
\end{equation}

\section{Benjamini-Hochberg and group false-discovery rate}
\label{app_fdr}
For a given method of selection of significant discoveries, let $D^F$ be the number of false positive and $D^T$ the number of true positive. The FDR $q$ is the mathematical expectation of the ratio between the number of false discoveries and the total number of discoveries $D^F/(D^F + D^T)$ (with the usual convention that $0/0=0$). The classic BH procedure was used to control the FDR. Let's first assume that the $p$ values have been sorted in ascending order, such that $p_{l}\leq p_{l+1}$. The BH procedure is built on an estimate $\hat{q}(p_l)$ of the false-positive rate, equal to $Lp_l/l$. The $p_l$ values are screened to find the largest $m$ such that $\hat{q}(p_{l})\leq \alpha$. If such an integer does not exist, there are no discoveries. Otherwise, all connections $l\leq m$ are considered as significant. 

\section{Generation of statistical parametric connectomes under the global null hypothesis}
\label{app_omnibus}
Let $\mathbf{Y}^{(s)}$ be the (subjects x connections) matrix of individual connectomes at scale $s$. A replication of the connectome matrix under the global null hypothesis $(\mathcal{G}_0)$ is generated by recomposing the linear mixture while excluding the $c$-th covariate of interest, tested by the model. Formally, let $\mathbf{X}_{\bar{c}}$ be the reduced model where the $c^\textrm{th}$ covariate has been removed from the (subjects x covariates) matrix $\mathbf{X}$. Let $\hat{\mathbf{B}}^{(s)}_{\bar{c}}$ be the ordinary least square estimate of the regression coefficients using the reduced model. Each permutation sample of the dataset is generated as described in \citep{Anderson2002}:
\begin{equation}
 \label{eq_samp_null}
 \mathbf{Y}^{(s,*)} = \mathbf{X}_{\bar{c}} \hat{\mathbf{B}}^{(s)}_{\bar{c}} + \hat{\mathbf{E}}^{(s,*)}.
\end{equation}
where $\hat{\mathbf{E}}^{(s,*)}$ is a replication of the residuals of the regression of the reduced model, with permuted rows (subjects). The GLM procedure is then implemented with the $\mathbf{Y}^{(s,*)}$ and the full model $\mathbf{X}$ to generate a replication $V^{(*)}_s$ of the volume of discoveries at scale $s$ under $(\mathcal{G}_0)$.
\par
Because the same dataset at voxel resolution is used to generate all the connectome datasets $(\mathbf{Y}^{(s)})_s$, the samples $V^{(*)}_s$ are not independent. In order to respect these dependencies, for any given replication, the same permutation of the subjects is used to generate to all of the $(\hat{\mathbf{E}}^{(s,*)})_s$. The replication of the total volume of discoveries $V^{(*)}$ is then simply the sum of $V^{(*)}_s$ for all $s$. This procedure is repeated $B$ times in order to generate $B$ replications $(V^{(*b)})_{b=1}^B$ of the total volume of discoveries under $(\mathcal{G}_0)$. The Monte-Carlo estimation of the probability to observe a greater total volume of discoveries under $(\mathcal{G}_0)$ than the actual total volume of discoveries $V$ generated on the original (non-permuted) dataset is then:
\begin{equation}
 \label{eq_pce_disc}
 \textrm{Pr}(V^{(*)}\geq V | \mathcal{G}_0) \doteq \#\left\{b=1,\dots,B|V^{(*b)}\geq V\right\}/B.
\end{equation}
where $\doteq$ means that the two terms are asymptotically equal as $B$ tends toward infinity. 

\clearpage
\pagebreak
\renewcommand{\thefigure}{S\arabic{figure}}
\renewcommand{\thetable}{S\arabic{table}}
\begin{center}
\emph{Supplementary Material {--} Multiscale statistical testing for connectome-wide association studies in fMRI}\\

\vspace{\baselineskip}Submitted to Neuroimage.\\

\vspace{\baselineskip}P. Bellec$^{1,2}$, Y. Benhajali$^{1,3}$, F. Carbonell$^{4}$, C. Dansereau$^{1,2}$, G. Albouy$^{1,5}$, M. Pelland$^{5}$, C. Craddock$^{6,7}$, O. Collignon$^{5}$, J. Doyon$^{1,5}$, E. Stip$^{8,9}$, P. Orban$^{1,8}$\\
\end{center}
$^1$Functional Neuroimaging Unit, Centre de Recherche de l'Institut Universitaire de G\'eriatrie de Montr\'eal\\
$^2$Department of Computer Science and Operations Research, University of Montreal, Montreal, Quebec, Canada\\
$^3$Department of Anthropology, University of Montreal, Montreal, Quebec, Canada\\
$^4$Biospective Incorporated, Montreal, Quebec, Canada\\
$^5$Department of Psychology, University of Montreal, Montreal, Quebec, Canada\\
$^6$Nathan Kline Institute for Psychiatric Research, Orangeburg, NY, United States of America\\
$^7$Center for the Developing Brain, Child Mind Institute, New York, NY, United States of America\\
$^8$Department of Psychiatry, University of Montreal, Montreal, Quebec, Canada\\
$^9$Centre Hospitalier de l'Universit\'e de Montr\'eal, Montreal, Quebec, Canada\\

For all questions regarding the paper, please address correspondence to Pierre Bellec, CRIUGM, 4545 Queen Mary, Montreal, QC, H3W 1W5, Canada. Email: pierre.bellec (at) criugm.qc.ca.\\

\pagebreak

\begin{landscape}
\begin{figure}[!ht]
\begin{center}
\includegraphics[width=\linewidth]{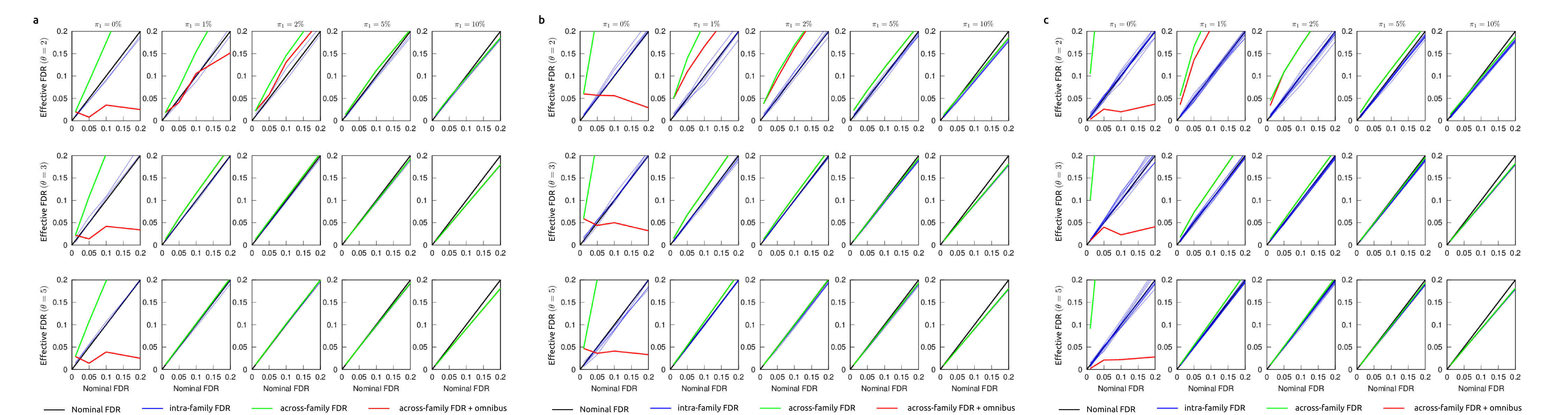}
\end{center}
\caption{
{\bf Nominal vs effective FDR on simulations with independent tests and variable $K$ ($L_k=1000$ for all $k$).} {The effective FDR is plotted against the nominal FDR within each family (blue plots), across all families (green plots) and across all families, combined with an omnibus test for rejection of the global null hypothesis (red plot). The expected (nominal) values are represented in black plots, corresponding to the four tested FDR levels: $0.01, 0.05, 0.1 ,0.2$. A variable number of families of tests were investigated, $K\in\{2,5,10\}$ for panels \textbf{a}, \textbf{b} and \textbf{c}, respectively. Each column corresponds to a certain proportion of non-null hypothesis per family $\pi_1$ ($0\%,1\%,2\%,5\%,10\%$), and each row corresponds to a different effect size $\theta$ ($2,3,5$), see text for details. Please note that in the presence of strong signal (large $\theta$ and/or $\pi_1$), the omnibus test is always rejected, and the green plot matches perfectly the red plot, which becomes invisible.} 
}
\label{fig_fdr_variable_K}
\end{figure}
\end{landscape}

\begin{landscape}
\begin{figure}[!ht]
\begin{center}
\includegraphics[width=\linewidth]{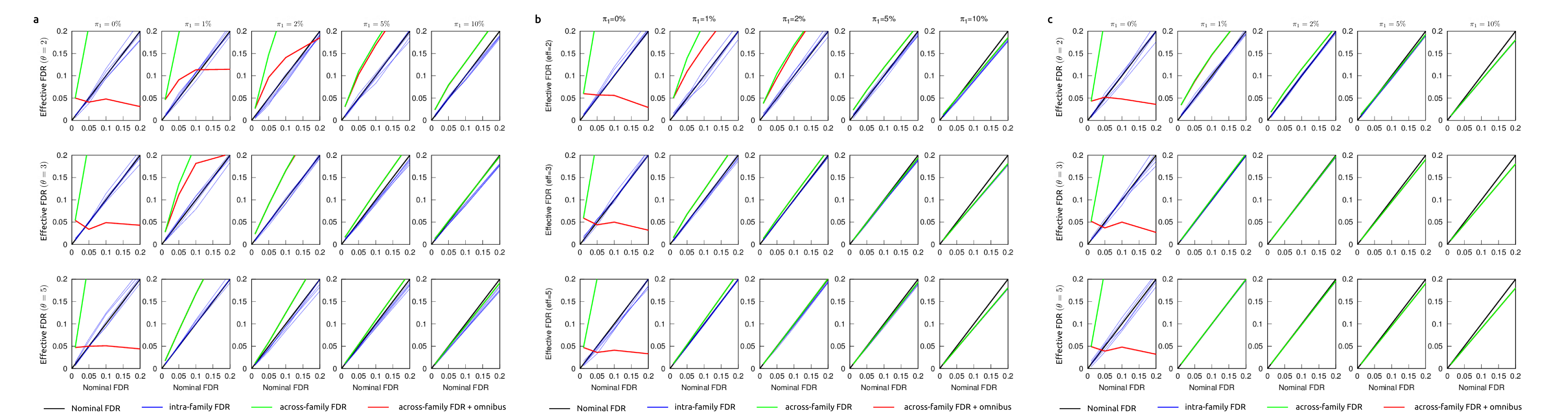}
\end{center}
\caption{
{\bf Nominal vs effective FDR on simulations with independent tests and variable $(L_k)_k$ ($K=5$).} {The effective FDR is plotted against the nominal FDR within each family (blue plots), across all families (green plots) and across all families, combined with an omnibus test for rejection of the global null hypothesis (red plot). The expected (nominal) values are represented in black plots, corresponding to the four tested FDR levels: $0.01, 0.05, 0.1 ,0.2$. A variable number of tests per family were investigated, $L_k \in \{100,1000,10000\}$ for panels \textbf{a}, \textbf{b} and \textbf{c}, respectively. Each column corresponds to a certain proportion of non-null hypothesis per family $\pi_1$ ($0\%,1\%,2\%,5\%,10\%$), and each row corresponds to a different effect size $\theta$ ($2,3,5$), see text for details. Please note that in the presence of strong signal (large $\theta$ and/or $\pi_1$), the omnibus test is always rejected, and the green plot matches perfectly the red plot, which becomes invisible.} 
}
\label{fig_fdr_variable_L}
\end{figure}
\end{landscape}

\begin{landscape}
\begin{figure}[!ht]
\begin{center}
\includegraphics[width=\linewidth]{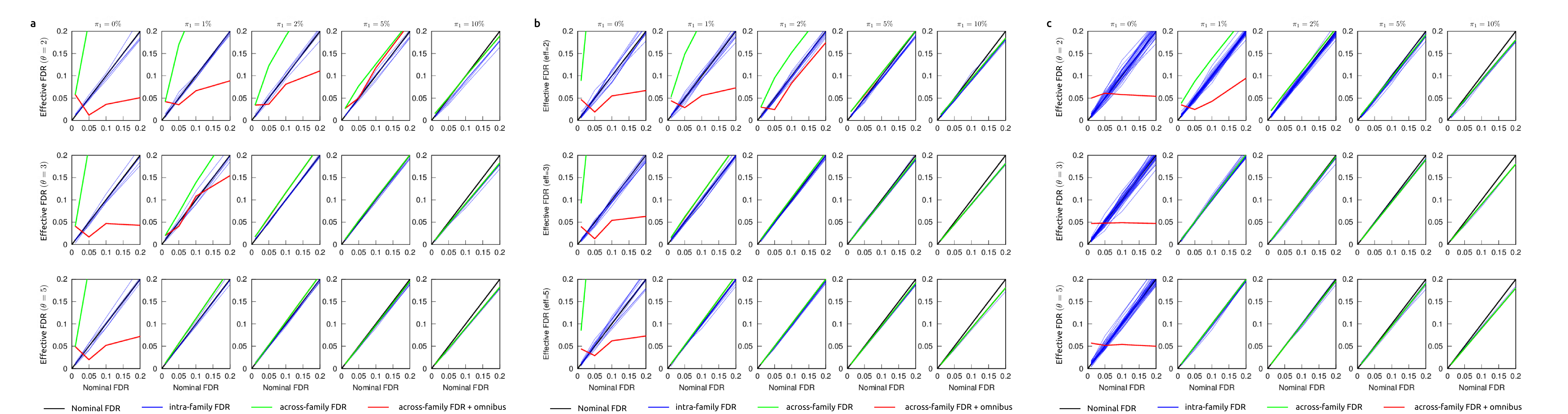}
\end{center}
\caption{
{\bf Nominal vs effective FDR on simulations with independent tests, $(L_k)_k$ corresponding to the number of connections associated with a regular grid of scales covering 10 to either 50 (a), 100 (b) or 300 (c) with a step of 10).} {The effective FDR is plotted against the nominal FDR within each family (blue plots), across all families (green plots) and across all families, combined with an omnibus test for rejection of the global null hypothesis (red plot). The expected (nominal) values are represented in black plots, corresponding to the four tested FDR levels: $0.01, 0.05, 0.1 ,0.2$. Each column corresponds to a certain proportion of non-null hypothesis per family $\pi_1$ ($0\%,1\%,2\%,5\%,10\%$), and each row corresponds to a different effect size $\theta$ ($2,3,5$), see text for details. Please note that in the presence of strong signal (large $\theta$ and/or $\pi_1$), the omnibus test is always rejected, and the green plot matches perfectly the red plot, which becomes invisible.} 
}
\label{fig_multiscale_variable_grid}
\end{figure}
\end{landscape}

\begin{figure}[!ht]
\begin{center}
\includegraphics[width=\linewidth]{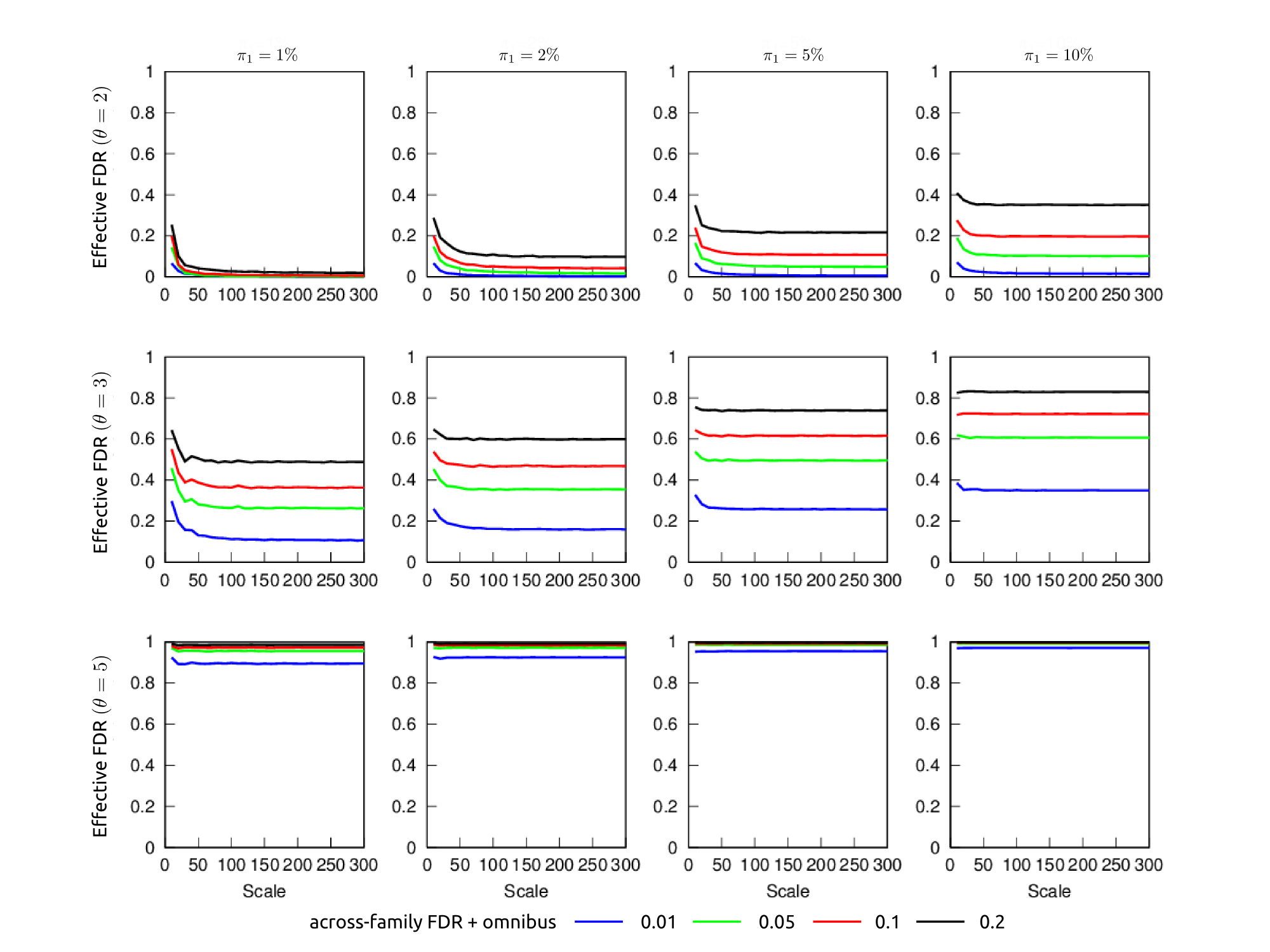}
\end{center}
\caption{
{\bf Sensitivity on simulations with independent tests ($K=30$, $L_k$ ranging from $55$ to $45150$, corresponding to the number of connections associated with a regular grid of scales covering 10 to 300 with a step of 10).} {The sensitivity is plotted as a function of scales at four tested (within-scale) FDR levels: $0.01, 0.05, 0.1 ,0.2$. A test is only considered as significant if in addition an omnibus test against the global null hypothesis across scales as been rejected at $p<0.05$. Each column corresponds to a certain proportion of non-null hypothesis per family $\pi_1$ ($0\%,1\%,2\%,5\%,10\%$), and each row corresponds to a different effect size $\theta$ ($2,3,5$), see text for details. } 
}
\label{fig_sens_multiscale_300_step_10}
\end{figure}
\pagebreak
\begin{figure}[!ht]
\begin{center}
\includegraphics[width=\linewidth]{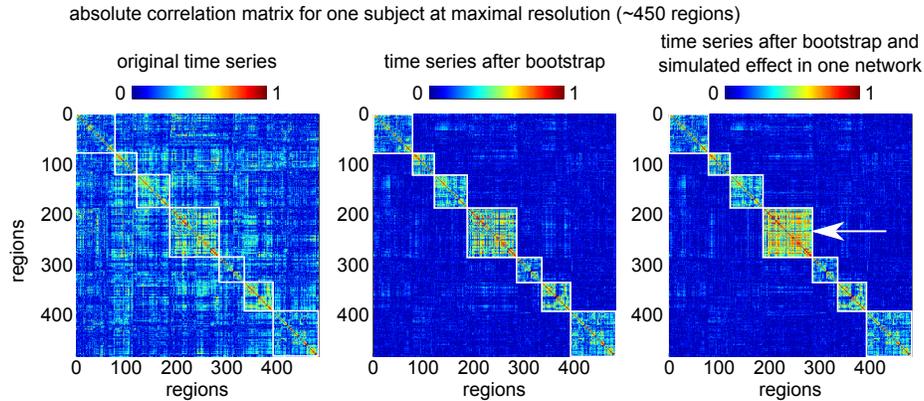}
\end{center}
\caption{
{\bf Data generation procedure for simulations.} {A hierarchical clustering applied on a group average connectome ($n=198$) was used to define partitions at multiple scales. A 7-cluster solution is presented as white squares outlining intra-clusters connections, superimposed on an individual connectome with rows/columns reordered based on the group hierarchy (left panel). A circular block bootstrap scheme is used to resample the original time series. Identical time blocks are used within each cluster, thus preserving intra-cluster connectivity. Independent time blocks are used between clusters, thus setting inter-parcel connectivity to zero (middle panel). A single simulated time series is added to all the regions belonging to one selected cluster, thus increasing the intra-parcel connectivity (right panel).} 
}
\label{fig_generation_simu}
\end{figure}

\begin{figure}[!ht]
\begin{center}
\includegraphics[width=\linewidth]{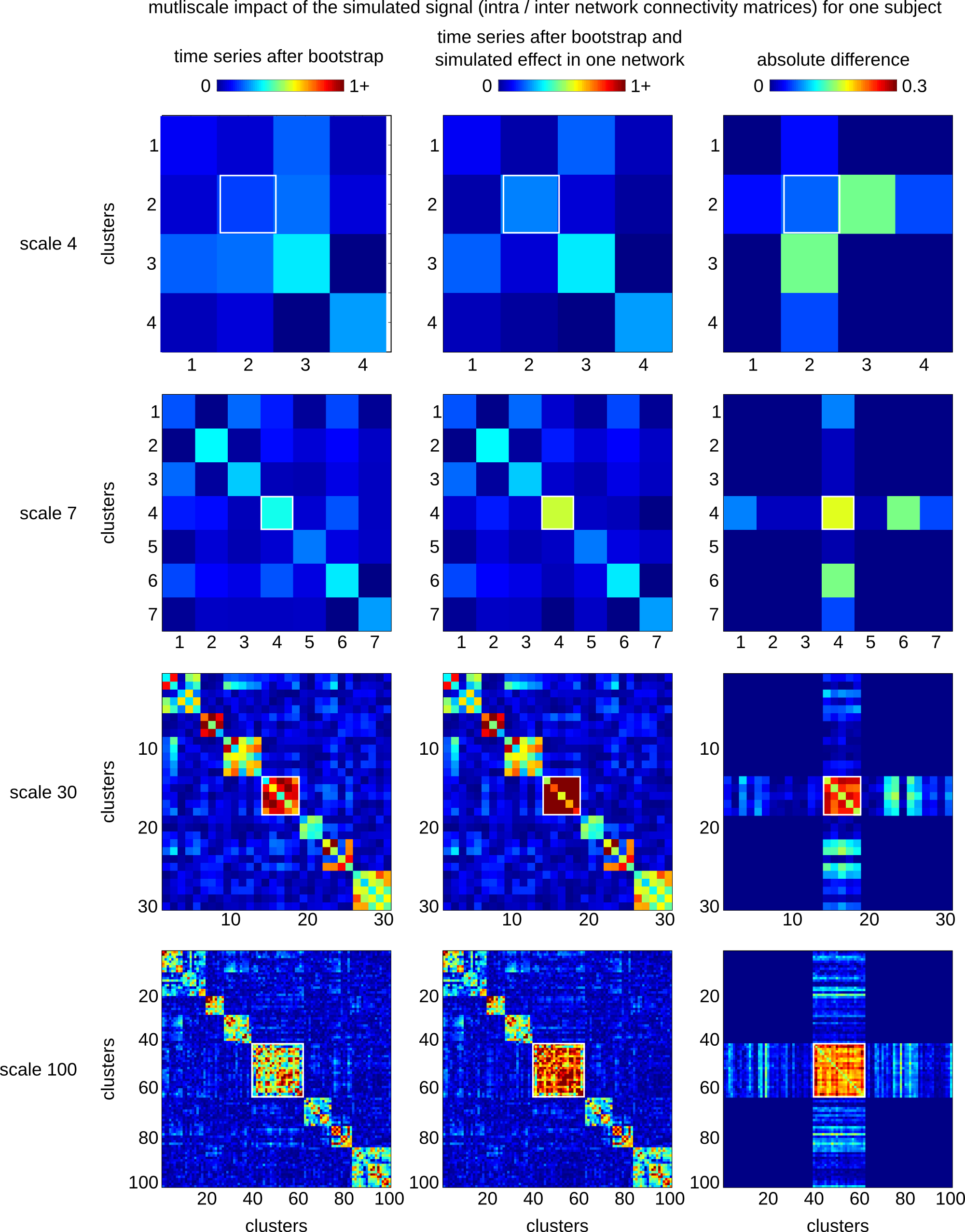}
\end{center}
\caption{
{\bf Impact of simulated changes on multiscale connectomes.} {The left column presents an individual connectome, after circular block resampling, at multiple scales (4, 7, 30, 100). The middle column shows the same connectome after a signal was injected in all regions belonging to one of the clusters at scale 7. The right column is the difference between the middle and the left column. Note how the main and only significant differences are concentrated in the connections that linked clusters that are either subclusters of the cluster of reference, or include the cluster of reference, as outlined by a white square. } 
}
\label{fig_generation_simu_multiscale}
\end{figure}

\begin{figure}[!ht]
\begin{center}
\includegraphics[width=\linewidth]{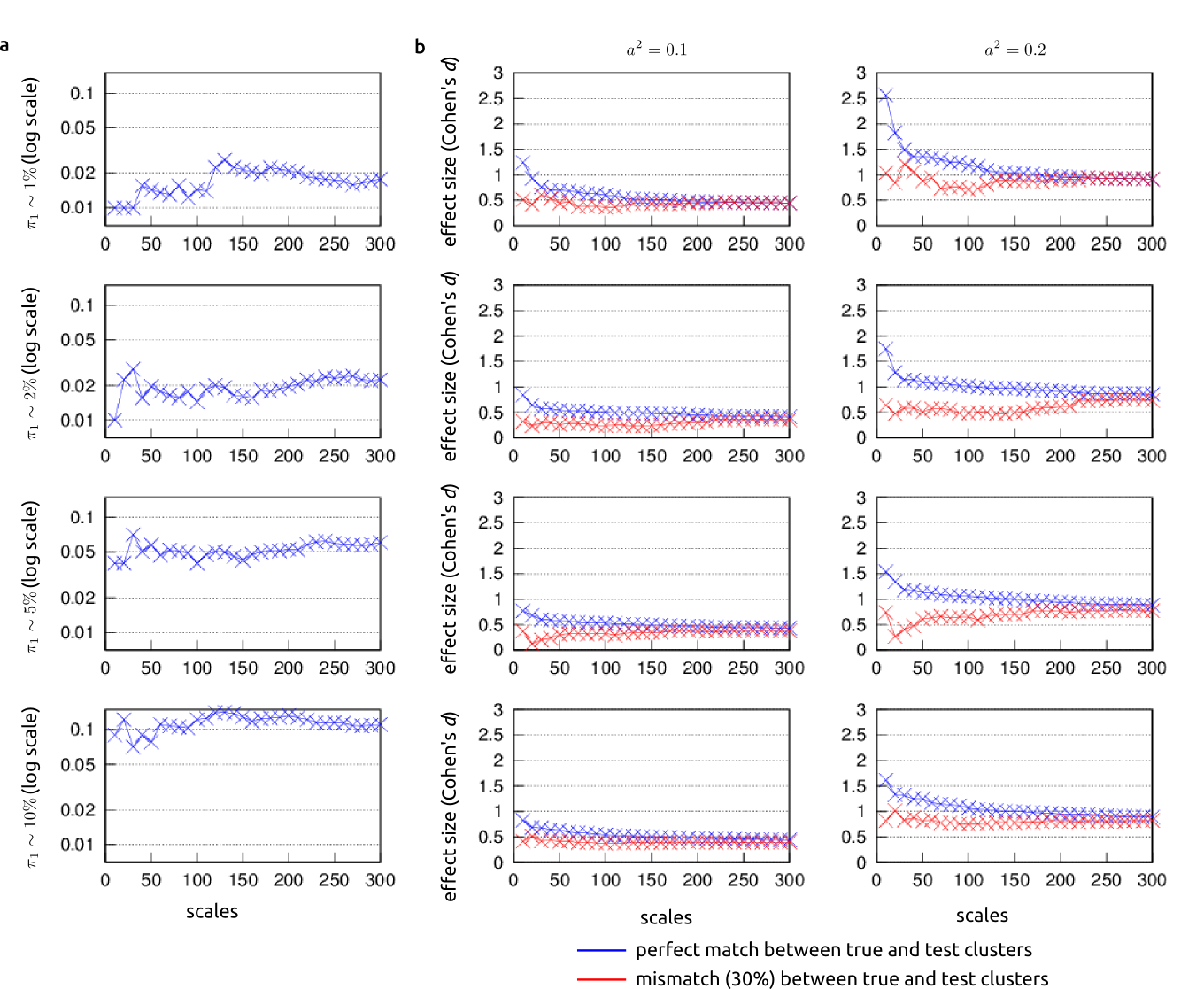}
\end{center}
\caption{
{\bf Percentage of true positive and effect size as a function of scale in the various simulation scenarios.} {Panel \textbf{a} shows the percentage of true positives in the simulation as a function of scale, for different choices of scale and cluster of reference. Panel \textbf{b} shows the average effect size over all true positives, for two choices of the parameter $a^2$ and the same choices of scale and cluster of references as in panel \textbf{a}. The effect size is plotted for simulations where the test and ground truth clusters exactly matched (blue curve) as well as for simulations where a perturbation of the reference cluster was applied (red curve).} 
}
\label{fig_eff_size_dep_multiscale_300_step_10}
\end{figure}

\begin{figure}[!ht]
\begin{center}
\includegraphics[width=\linewidth]{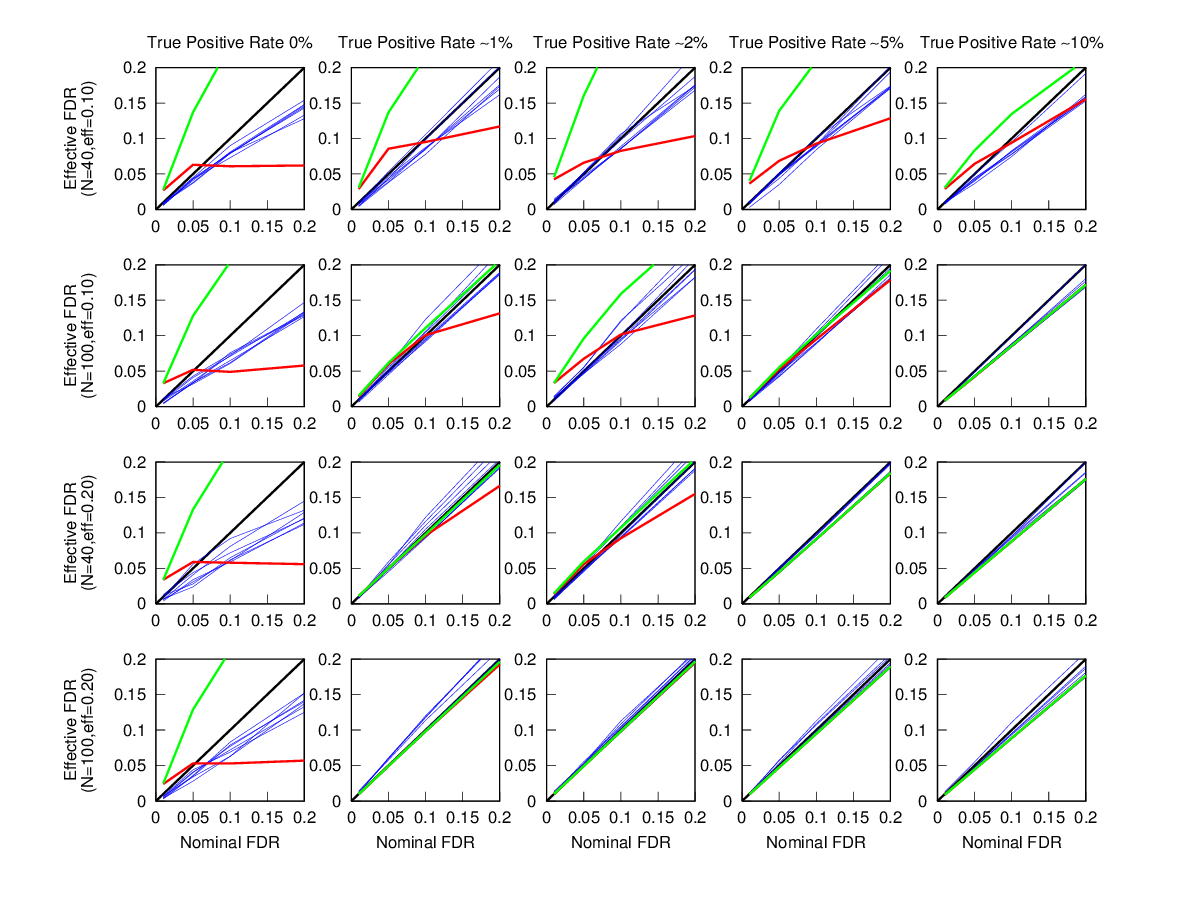}
\end{center}
\caption{
{\bf Nominal vs effective FDR on simulations with dependent tests ($K=7$, $L_k$ in $(28,136,325,1540,6555,19900,53956)$, corresponding to the number of connections associated with the scales selected by MSTEPS on the SCHIZO dataset).} {The effective FDR is plotted against the nominal FDR within each family (blue plots), across all families (green plots) and across all families, combined with an omnibus test for rejection of the global null hypothesis (red plot). The expected (nominal) values are represented in black plots, corresponding to the four tested FDR levels: $0.01, 0.05, 0.1 ,0.2$. Each column corresponds to a certain proportion of non-null hypothesis per family $\pi_1$ (about $0\%,1\%,2\%,5\%,10\%$), and each row corresponds to a combination of a different effect size, $a^2$ in $\{0.1,0.2\}$, and number of subjects, $N$ in $\{20,50\}$, see text for details. Please note that in the presence of strong signal (large $a^2$,  $N$ and/or $\pi_1$), the omnibus test is always rejected, and the green plot matches perfectly the red plot, which becomes invisible.} 
}
\label{fig_simu_dep_msteps_schizo}
\end{figure}

\begin{table}[!ht]
 \begin{center}
  \begin{tabular}{llllllllll}
   MOTOR & $K$ & & 10 & 20 & 40 & 90 & 170 & 270\\
         & $L$ & & 7  & 16 & 36 & 72 & 153 & 297\\
         & $M$ & & 7  & 14 & 35 & 72 & 154 & 308\\
   \hline
   BLIND & $K$ & & 10 & 20 & 50 & 90 & 180 & 280\\
         & $L$ & & 7  & 18 & 40 & 81 & 162 & 308\\
         & $M$ & & 7  & 16 & 40 & 77 & 165 & 313\\
   \hline
   SCHIZO & $K$ & 10 & 20 & 30 & 60 & 120 & 210 & 270\\
          & $L$ & 7  & 14 & 27 & 54 & 108 & 189 & 324\\
          & $M$ & 7  & 16 & 25 & 55 & 114 & 199 & 328\\
  \end{tabular}
 \end{center}
 \caption{Summary of the scales selected by MSTEPS on the real data samples. Three parameters were selected by MSTEPS at each scale: $K$ was the number of individual clusters, identical for all subjects; $L$ was the number of group clusters, used to compute the group stability matrix in BASC; $M$ was the number of final clusters for the group consensus cluster analysis. The effective number of clusters in the group template is $M$. The two other parameters ($K$ and $L$) are used in intermediate computation of the multi-level BASC. See details of the parameters of the MSTEPS procedure in the main text. }
\label{tab_msteps}
\end{table}

\begin{figure}[!ht]
\begin{center}
\includegraphics[width=\linewidth]{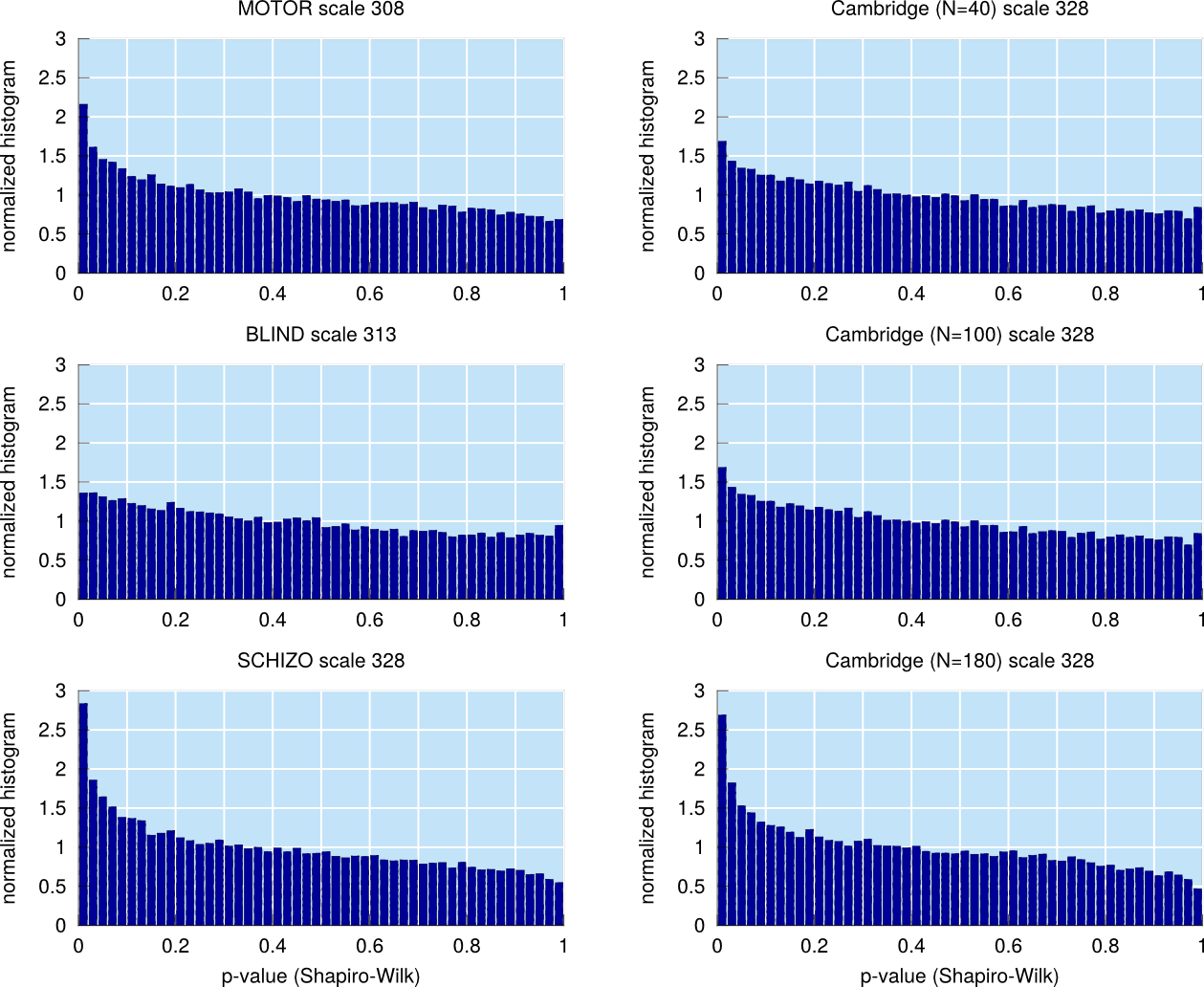}
\end{center}
\caption{
{\bf Test on the normality of the distribution of residuals in a GLM-connectome analysis.} {Distribution (normalized histogram) of $p$-values derived using the Shapiro-Wilk parametric hypothesis test of composite normality across all connections in various GLM-connectome analyses, at the highest scale (300+) selected by MSTEPS. Note that for Cambridge, random groups of equal size were compared, for different sample sizes. In the absence of a deviation of Gaussian resiudals, the histogram would be flat (equal to 1). Although some trend towards a departure can clearly be seen, with an excess of small $p$-values in the distribution, no $p$-value reaches significance after correction of multiple comparisong using the FDR-BH procedure at $q<0.05$.} 
}
\label{fig_test_gaussian}
\end{figure}

\begin{figure}[!ht]
\begin{center}
\includegraphics[width=\linewidth]{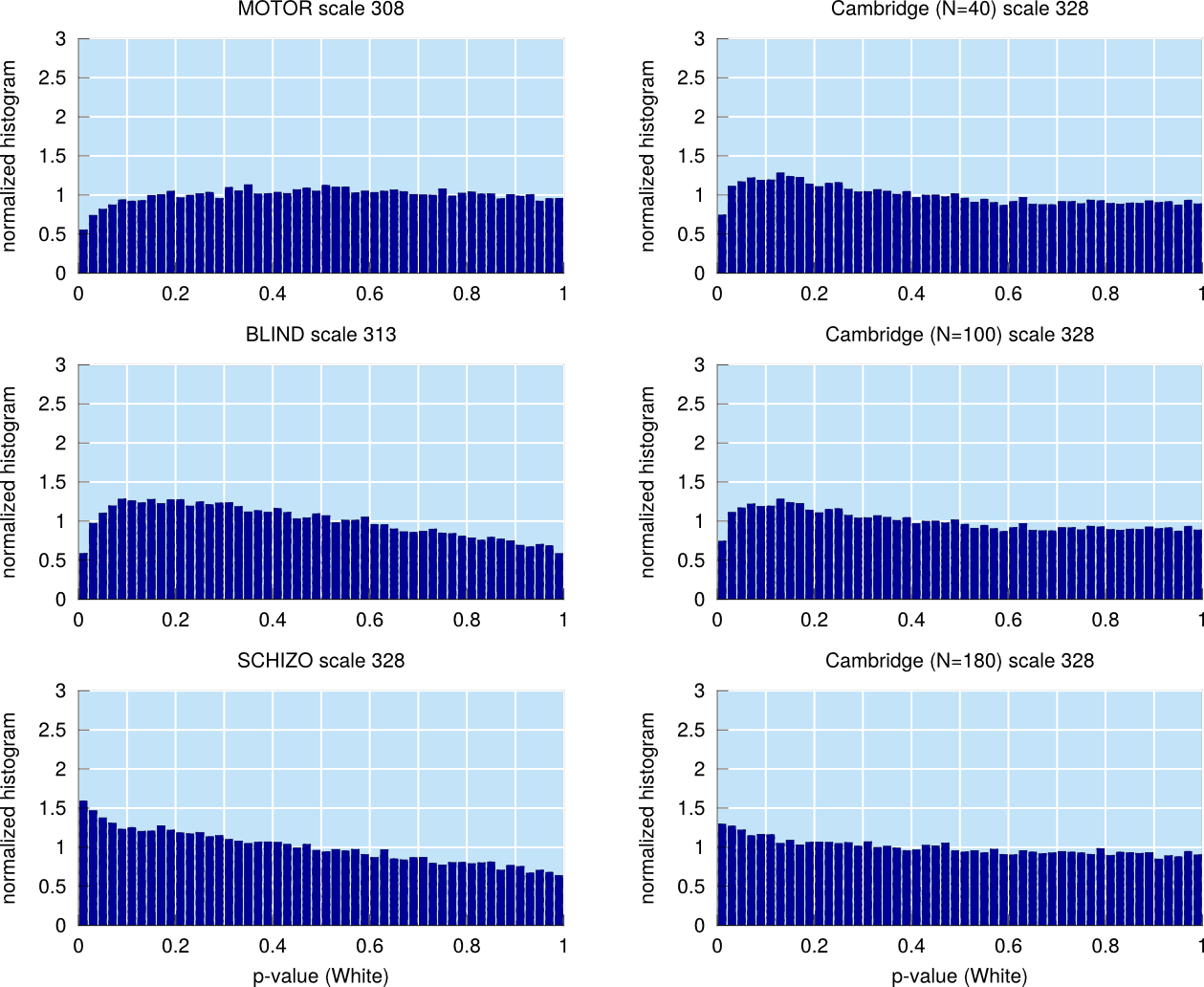}
\end{center}
\caption{
{\bf Test on the homoscedasticity of residuals in a GLM-connectome analysis.} {Distribution (normalized histogram) of $p$-values derived using White's test of homoscedastic residuals across all connections in various GLM-connectome analyses, at the highest scale (300+) selected by MSTEPS. Note that for Cambridge, random groups of equal size were compared, for different sample sizes. Due to the random nature of grouping, the residuals in the Cambridge contrasts are homoscedastic, and the expected histogram is flat (equal to 1). The analysis on the BLIND, MOTOR and SCHIZO datasets resulted in histograms similar to those observed on the Cambridge sample, for different sample sizes. No $p$-value reaches significance after correction of multiple comparisong using the FDR-BH procedure at $q<0.05$.} 
}
\label{fig_test_hetero}
\end{figure}

\end{document}